\pdfoutput=1

\documentclass[12pt]{article} 
\usepackage{hyperref}
\usepackage{url}

\usepackage[parfill]{parskip}

\usepackage{verbatim}
\usepackage{amsmath}
\usepackage{amsthm}
\usepackage{amssymb}
\usepackage{bm}

\usepackage{lscape}
\usepackage{graphicx}
\usepackage{wrapfig}
\usepackage{verbatim}
\usepackage{helvet}
\usepackage{float}
\usepackage{caption}
\usepackage[caption=false,font=footnotesize]{subfig}
\usepackage{calc}
\usepackage{xspace}
\usepackage{pifont}
\usepackage{multirow}
\usepackage{hyperref}
\usepackage{placeins}
\usepackage[normalem]{ulem}
\usepackage{movie15}
\usepackage{stackengine}

\usepackage[top=0.75in,bottom=1in,left=1in,right=1in,foot=0.5in,]{geometry}

\def\proof{\noindent{\bf Proof:} }
\def\qed{\hfill{$\Box$} \\}

\def\real{{\mathbb R}}

\def\hilbert{{\mathcal H}}

\newtheorem{theorem}{Theorem}

\newtheorem{lemma}{Lemma}

\newtheorem{definition}{Definition}
\newtheorem{remark}{Remark}

\newtheorem{algorithm}{Algorithm}

\usepackage{color}

\def\videoinline{true}

\title{Sparse Functional Identification of Complex Cells from Spike Times and the Decoding of Visual Stimuli}

\author{
Aurel A. Lazar \and Nikul H. Ukani \and Yiyin Zhou\thanks{The authors' names are listed in alphabetical order.}\\
Department of Electrical Engineering\\
Columbia University\\
New York, NY 10027 \\
\texttt{aurel, nikul, yiyin@ee.columbia.edu} \\
}

\begin{document}

\maketitle

\begin{abstract}

We investigate the sparse functional identification of
complex cells and the decoding of visual stimuli encoded by an ensemble of complex cells.
The reconstruction algorithm of both temporal and
spatio-temporal stimuli is formulated as a rank minimization problem that significantly reduces the number of sampling
measurements (spikes) required for decoding.
We also establish the duality between sparse decoding and functional identification, and provide algorithms for identification of low-rank dendritic stimulus processors.
The duality enables us to efficiently evaluate our functional identification algorithms by reconstructing novel stimuli in the input space.
Finally, we demonstrate that our identification algorithms
substantially outperform the generalized quadratic model, the non-linear input model and the widely used spike-triggered covariance  algorithm.

\noindent
\end{abstract}

\newpage
\tableofcontents
\newpage

\section{Introduction}
\label{sec:intro}

\begin{figure}[h!p]
\centering
\includegraphics[width=0.7\textwidth]{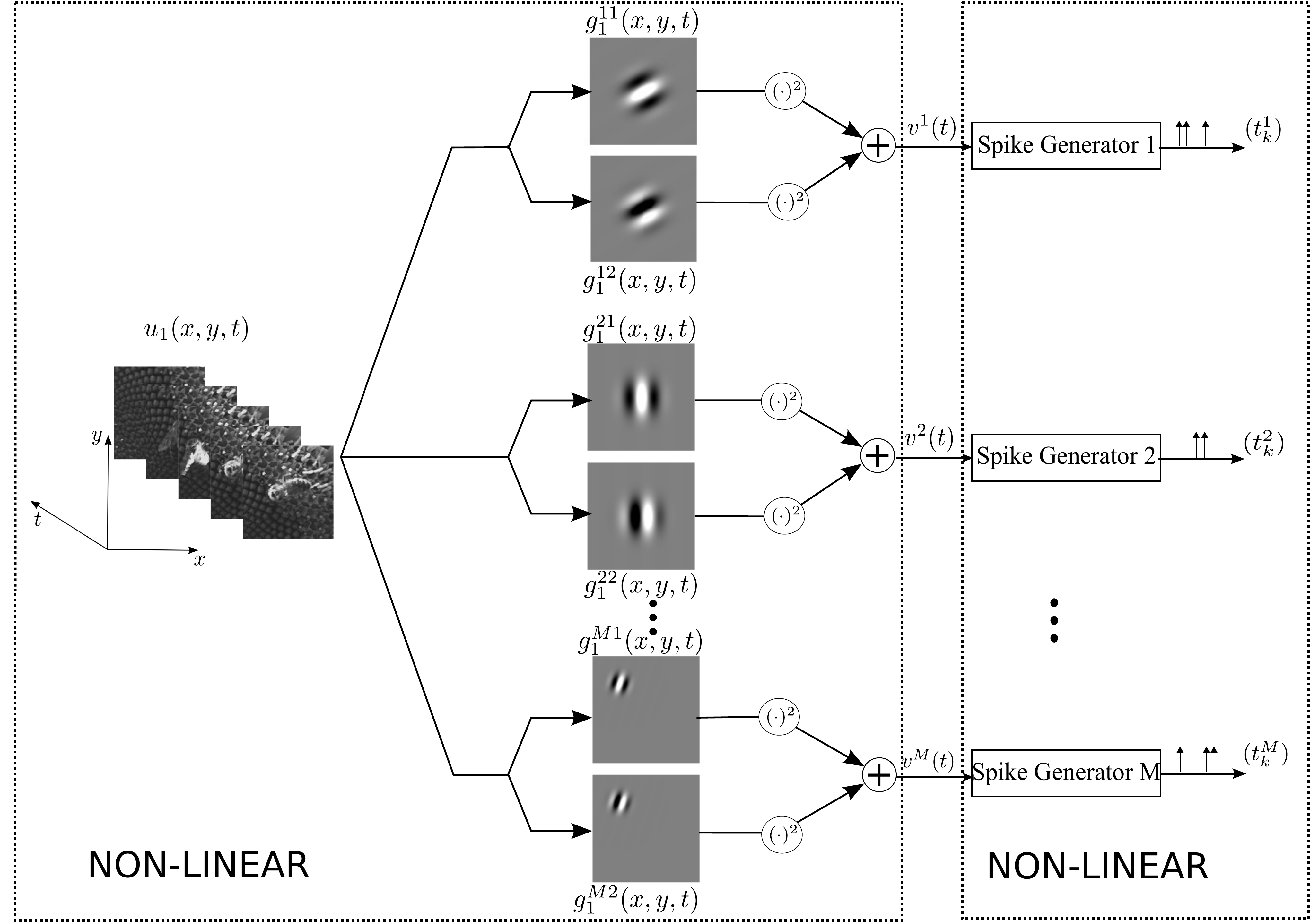}
\caption{A neural circuit consisting of a population of complex cells.}
\label{fig:complex}
\end{figure}

It is widely accepted that the early mammalian visual system employs a series of neural circuits to extract
elementary visual features, such as edges and motion \cite{HW1962, BAL1965}.
Feature extraction capabilities of simple and complex cells arising in the primary visual cortex (V1) have been extensively investigated. 
Each simple cell consists of a linear receptive field cascaded with a highly-nonlinear spike generator.
Similar to simple cells, complex cells in V1 are selective to oriented edges/lines
over a spatially restricted region of the visual field \cite{HW1962}.
While simple cells respond maximally to a particular
phase of the edge,
complex cells are largely phase invariant \cite{RH2002}. Therefore, the 
receptive fields of complex cells cannot be simply mapped into excitatory and inhibitory regions \cite{HW1962}.
Receptive fields of simple cells can be modeled
as linear Gabor filters while processing in complex cells can be modeled
with a quadrature pair of Gabor filters followed by 
squaring \cite{AB1985}. Neural circuits comprising complex cells constitute a highly nonlinear circuit as illustrated in 
Figure~\ref{fig:complex}.
		
Under the modeling framework of Time Encoding Machines (TEMs) \cite{LP2011,LPZ2010,LAY14}, it has been shown that 
decoding of stimuli and functional identification of linear receptive fields 
of simple cells are dual to each other \cite{LAS13,LSZ2015}.
This led to mathematically rigorous
identification algorithms for identifying linear receptive fields of simple cells \cite{LAY16}.
By modeling the nonlinear processing in complex cells
as Volterra Dendritic Stimulus Processors (DSPs) \cite{LAS15,LAY14a}, the 
representation of stimuli encoded by spike times generated by neural circuits with complex cells was also exhaustively analyzed.
Functional identification of a complex cell DSP was possible again thanks to the demonstrated duality between decoding
and functional identification.
While these theoretical methods exhibit deep structural properties, they have been shown to be tractable only for  decoding and functional identification problems of
small dimensions.
In their current form they are not tractable due to the ``curse of dimensionality'' \cite{Marmarelis2004}.

The non-linear
transformations taking place in the DSP of complex cells
lead to loss of phase information.
With this in mind, we formulate the reconstruction of
stimuli encoded with complex cells as a phase retrieval problem \cite{CSV2011} and, in search of tractable algorithms,   
utilize recent developments in optimization theory of
low-rank matrices \cite{CR09, RFP10, CSV2011}.
By applying such methods, we develop algorithms that are highly effective
in decoding visual stimuli encoded by complex cells.
As will be detailed in the next sections, the complex cells, as defined in this paper, have DSP kernels that are low-rank
and include the ones shown in Figure \ref{fig:complex} as a particular case.

After demonstrating that the decoding of visual stimuli becomes tractable, we propose sparse algorithms for functionally identifying the DSPs of complex cells using the spike times they generate. 
The sparse identification algorithms are based on the key observation 
that functional identification
can be viewed as the dual problem of decoding stimuli that are
encoded by an ensemble of complex cells. While the generalization of the duality results from simple cells to complex
cells was already given in \cite{LAS15},
we show in this paper that these results remain valid under
the assumption of sparsity, that is, for the case of low-rank DSP kernels.
This significantly reduces the time of stimulus presentation that is needed in the identification process.
The sparse duality result also enables us to evaluate the identified
circuits in the input space.
We achieve the latter by computing the mean square error or signal-to-noise (SNR) of novel stimuli decoded using the identified circuits \cite{LAS13,LSZ2015}.

This paper is organized as follows. In Section~\ref{sec:volterra_tem}, 
we first introduce the modeling of encoding of temporal
stimuli with complex cells.
We provide a detailed review of decoding of stimuli and the functional identification of complex cells,
and point out the current algorithmic limitations.
In Section~\ref{sec:rankmin}, we provide sparse decoding algorithms that achieve high accuracy and are algorithmically 
tractable.
We then explicate the dual relationship
between sparse functional identification and decoding and
provide examples for the identification of low-rank, temporal DSP kernels of complex cells.
In Section~\ref{sec:extension}, we extend sparse decoding
methodology to spatio-temporal stimuli and functional identification of spatio-temporal complex cells.
Using novel stimuli, we provide evaluation examples of the
identification algorithms in the input space as well as 
comparisons to other state-of-the-art methods.
Finally, we conclude in Section~\ref{sec:conc}
and suggest how the approach advanced in this paper can be applied beyond complex cells.

\section{Neural Circuits with Complex Cells:\\ Encoding, Decoding and Functional Identification}
\label{sec:volterra_tem}

In this section, we model the encoding of temporal stimuli by a neural circuit
consisting of neurons akin to complex cells.
We start by modeling the space of temporal stimuli in Section~\ref{sec:temp_stim_model}.
In Section~\ref{sec:temp_encode}, the model of encoding is formally described.
In Section~\ref{sec:VTDM}, we proceed to present a reconstruction algorithm for decoding temporal stimuli
encoded by the neural circuit.
A method for functional identification of neurons constituting the neural circuit is provided
in Section~\ref{sec:volterra_cim}.
The reconstruction algorithm and the
functional identification algorithm discussed in this section are based on \cite{LAS15}.

\subsection{Modeling Temporal Stimuli}
\label{sec:temp_stim_model}

We model the temporal varying stimuli
$u_1=u_1(t)$, $t \in \mathbb{D}$, to be real-valued
elements of the space of trigonometric polynomials \cite{LPZ2010}.
The choice of the space of the trigonometric polynomials has, as
we will see, substantial computational advantages.
\begin{definition}
The space of trigonometric polynomials 
$\hilbert_1$ is the Hilbert space of
complex-valued functions
\begin{equation}
u_1(t) = \sum_{l_t=-L_t}^{L_t} c_{l_t}e_{l_t}(t),
\label{eq:uform}
\end{equation}
over the domain $\mathbb{D} =  [0, S_t]$, where
\[
e_{l_t}(t) =  \frac{1}{\sqrt{S_t}}\operatorname{exp}\left(\frac{jl_t\Omega_t}{L_t}t \right).
\]
Here $\Omega_t$ denotes the bandwidth,
and $L_t$ the order of the space.
Stimuli $u_1\in\hilbert_1$ are extended to be periodic over $\real$ with period
$S_t = \frac{2\pi L_t}{\Omega_t}$.
\end{definition}

$\hilbert_1$ is a Reproducing Kernel Hilbert Space (RKHS)
\cite{BTA04} with
reproducing kernel (RK)
\vspace{-0.075in}
\begin{equation}
K_1(t;t') = \sum_{l_t=-L_t}^{L_t}e_{l_t}(t-t').
\end{equation}

We denote the dimension of $\hilbert_1$ by $dim(\hilbert_1)$ and 
$dim(\hilbert_1) = 2L_t+1$.

\begin{definition}
The tensor product space $\hilbert_2 = \hilbert_1 \otimes \hilbert_1$ is a Hilbert space of complex-valued functions
\begin{equation}
u_2(t_1;t_2) =\!\!\! \sum_{l_{t_1}=-L_t}^{L_t}\sum_{l_{t_2}=-L_t}^{L_t} d_{l_{t_1} l_{t_2}} e_{l_{t_1}}(t_1) \cdot e_{l_{t_2}}(t_2)
\label{eq:u2form}
\end{equation}
over the domain $\mathbb{D}^2$.
\end{definition}

$\hilbert_2$ is an RKHS with reproducing kernel
\vspace{-0.1in}
\[
K_2(t_1,t_2;t'_1,t'_2) = \sum_{l_{t_1}=-L_t}^{L_t}\sum_{l_{t_2}=-L_t}^{L_t}e_{l_{t_1}}(t_1-t'_1) \cdot e_{l_{t_2}}(t_2-t'_2).
\]

Note that $dim(\hilbert_2) = dim(\hilbert_1)^2$.

\subsection{Encoding of Temporal Stimuli by a Population of Complex Cells}
\label{sec:temp_encode}

\begin{figure}[t] 
\centering
\subfloat[]{ \label{fig:cc_square} \includegraphics[width=0.45\textwidth]{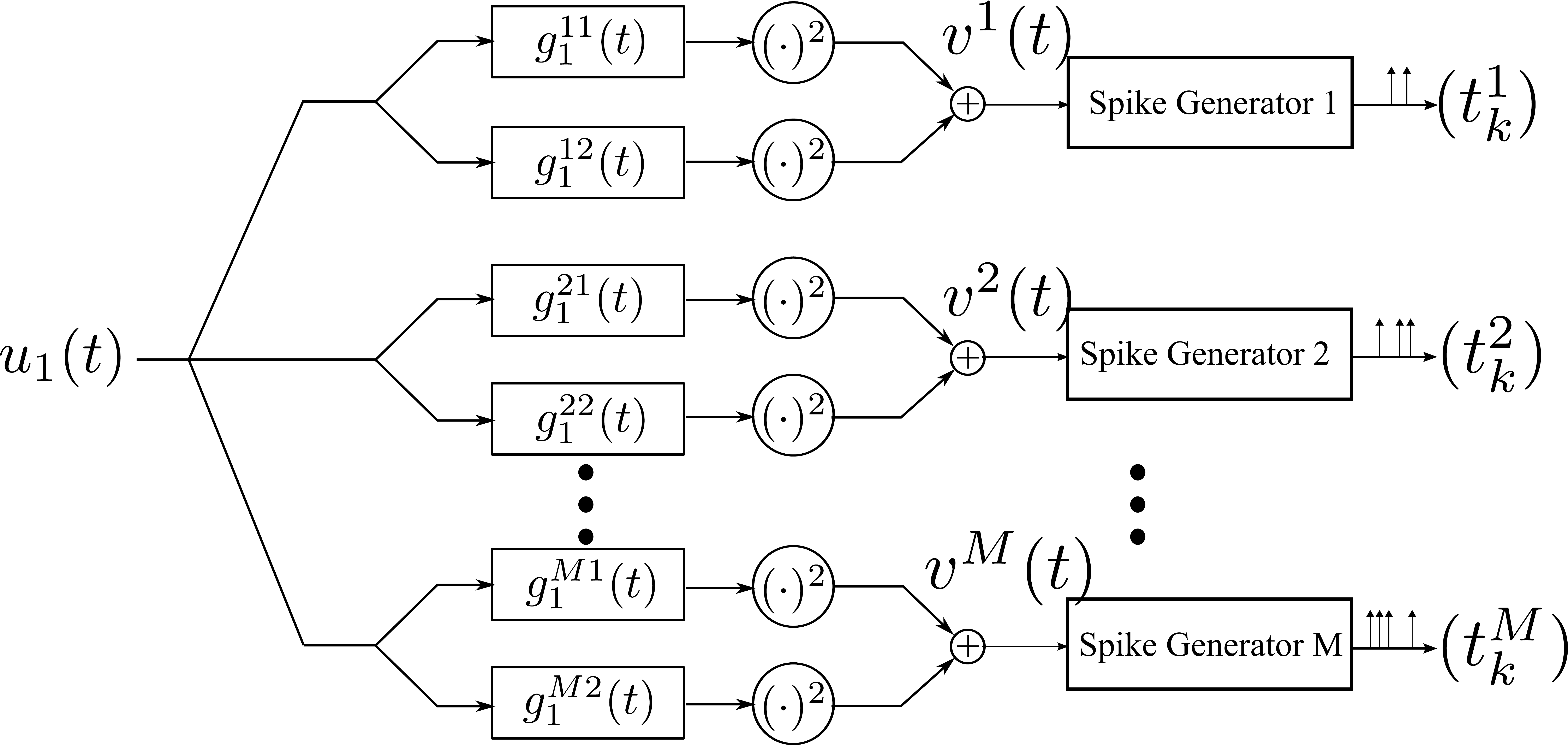}} ~~
\subfloat[]{ \label{fig:cc_square2} \includegraphics[width=0.45\textwidth]{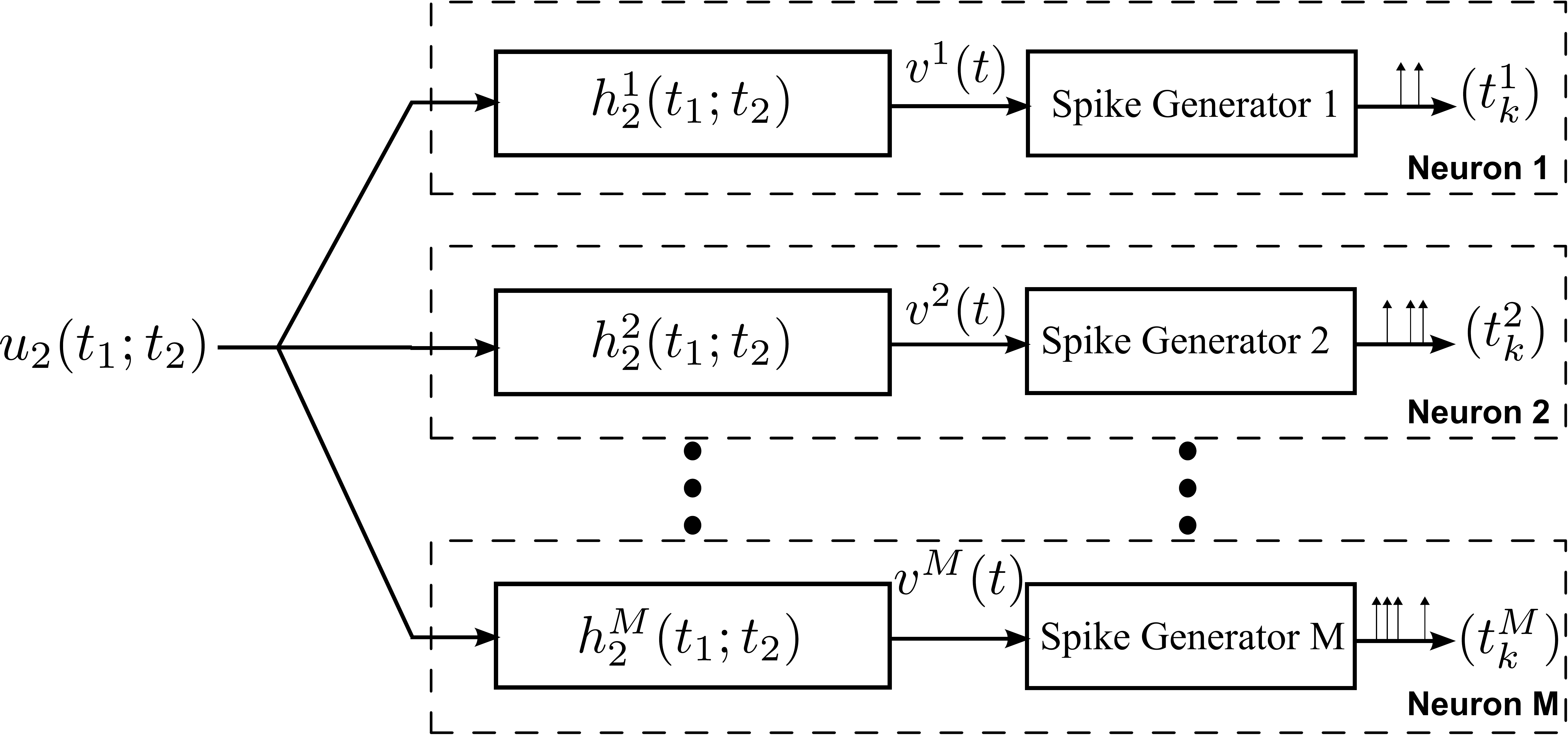}}
\caption{The encoding of temporal stimuli by a neural circuit modeling an ensemble of neurons akin to complex cells.
(a) The $i^{\text{th}}$ neuron in the model processes the input $u_1(t)$ by two parallel linear filters with impulse responses
$g^{i1}_1(t)$ and $g^{i2}_1(t)$, respectively, followed by squaring. The outputs are summed and then fed into a spike generator.
(b) An equivalent representation of the encoding circuit in which the DSPs are represented as second-order Volterra kernels.}
\label{fig:volterra_tem}
\end{figure}

We consider a neural circuit consisting of  $M$ neurons 
as shown in Figure~\ref{fig:cc_square}.
For the $i^{\text{th}}$ neuron, input stimulus $u_1(t)$ ($u_1\in \hilbert_1$)
is first processed by two linear filters with impulse responses $g^{i1}_1(t)$ and $g^{i2}_1(t)$,
the outputs of which are individually squared and then summed together.
These processing elements are
integral part of the DSP of neuron $i$ \cite{LAS15,LAY14a}.
The output of the DSP $i$, denoted by $v^i(t)$, is then
fed into the Biological Spike Generator (BSG) of neuron $i$.
The BSG $i$ encodes the output of DSP $i$ into 
the spike train $(t^i_k)_{k\in\mathbb{I}^i}$.
Here $\mathbb{I}^i$ is the spike train index set of neuron $i$.
We notice the similarity between the overall structure of neural circuits in Figure~\ref{fig:cc_square} and Figure~\ref{fig:complex}.
In what follows, we refer to the neurons in the neural circuit in Figure~\ref{fig:cc_square} as complex cells.

The output of the DSP of the $i^{th}$ neuron in Figure~\ref{fig:cc_square} amounts to
\begin{equation}
v^i(t) = \left[\int_{\mathbb{D}} g^{i1}_1(t-s_1) u_1(s_1) ds_1 \right]^2 +
\left[\int_{\mathbb{D}} g^{i2}_1(t-s_2) u_1(s_2) ds_2 \right]^2,
\label{eq:cc}
\end{equation}
for all $i=1,2,\cdots,M$.

With 
\begin{equation}
h^i_2(t_1;t_2) = g^{i1}_1(t_1)g^{i1}_1(t_2)+g^{i2}_1(t_1)g^{i2}_1(t_2),
\label{eq:gtoh}
\end{equation}
\eqref{eq:cc} can be  rewritten as  
\begin{equation}
v^i(t) = \int_{\mathbb{D}^2} h^i_2(t-s_1;t-s_2) u_1(s_1) u_1(s_2) ds_1 ds_2 ,
\label{eq:volterra_output}
\end{equation}
where $\mathbb{D}^2$ denotes the Cartesian product of the domain $\mathbb{D}$
of $u_1$ and $ h^i_2(t_1;t_2)$ is interpreted as a second-order Volterra kernel \cite{Rugh1981}. 
We assume that
$h_2^i(t_1;t_2)$ is 
real, bounded-input bounded-output (BIBO) stable, causal and of finite memory.
The I/O of the neural circuit shown in Figure~\ref{fig:cc_square} can be equivalently outlined as in Figure~\ref{fig:cc_square2},
in which each neuron processes the input $u_1(t)$ nonlinearly by a second order kernel $h^i_2(t_1;t_2)$ followed by a BSG.

\begin{remark}
Note that the BSG models the spike generation mechanism  of
the axon hillock of a biological
neuron, whereas the DSP is an equivalent model of processing
of the stimuli by a sophisticated neural network that proceeds the spike generation. 
Therefore, stimulus processing and the spike generation mechanism are naturally separated in the neuron model considered here.
\end{remark}

For simplicity, we will use the Integrate-and-Fire (IAF) neuron model as the spike generation mechanism (see, e.g., \cite{LP2011}).
Note that, the algorithms described here can also be employed with other spike generators such as the Hodgkin-Huxley, Morris-Lecar and Izhikevic neuron models \cite{HH1952, ML1981, Izhikevich03, AKL11, LAY14a}.
The integration constant, bias and threshold of the IAF neuron $i=1,2,\cdots,M$, are denoted by $\kappa^i$, $b^i$ and $\delta^i$, respectively.
The t-transform of the $i$-th IAF neuron is given by \cite{LP2011,LPZ2010,LAY14}
\begin{equation}
\int_{t^i_k}^{t^i_{k+1}} v^i(t)dt = \kappa^i\delta^i - b^i(t^i_{k+1}-t^i_{k}).
\label{eq:iaf}
\end{equation}

\begin{lemma}
The encoding of the temporal stimulus $u_1\in\hilbert_1$ into the spike train sequence $(t_k^i), k\in\mathbb{I}^i$, $i=1,2,...,M$, by a neural circuit with complex cells
is given in functional form by
\begin{equation}
\mathcal{T}^i_k u_2 = q^i_k, k\in\mathbb{I}^i, i = 1,\cdots, M,
\label{eq:t-trans}
\end{equation}
where $M$ is the total number of neurons, $n_i+1$ is the number of spikes generated by neuron $i$ and 
$\mathcal{T}^i_k: \hilbert_2 \rightarrow \real$,
are bounded linear functionals defined by
\begin{equation}
\mathcal{T}^i_k u_2 = \int_{t^i_k}^{t^i_{k+1}} \int_{\mathbb{D}^2} h^i_2 (t-s_1;t-s_2) u_2(s_1; s_2) ds_1 ds_2 dt,
\label{eq:operatorT}
\end{equation}
with 
$u_2(t_1;t_2) = u_1(t_1)u_1(t_2)$.
Finally, $q^i_k = \kappa^i\delta^i - b^i (t^i_{k+1}-t^i_{k})$.
\label{lemma:linear-t-transform}
\end{lemma}
\proof 
The relationship \eqref{eq:t-trans} follows by replacing
the functional form of $v^i(t)$ given in \eqref{eq:volterra_output} in
equation \eqref{eq:iaf} above. \qed
\begin{remark}
\label{rem:gen_sampling}
$u_2(t_1,t_2)=u_1(t_1)\cdot u_1(t_2)$ can be interpreted as a nonlinear map of the
stimulus $u_1$ into $u_2$ defined in a higher dimensional space.
The operation performed by the
second order Volterra kernel on $u_2$ in \eqref{eq:operatorT} is linear.
Thus, \eqref{eq:t-trans} shows that the encoding of temporal 
stimuli can be viewed as generalized sampling \cite{LAS15}.
\end{remark}

\subsection{Decoding of Temporal Stimuli Encoded by a Population of Complex Cells}
\label{sec:VTDM}

Assuming that the spike times $(t_k^i), k\in \mathbb{I}^i, i=1,2,...,M$, are known, by Lemma~\ref{lemma:linear-t-transform},
the neural circuit with complex cells encodes the stimulus
via a set of linear functionals acting
on $u_2$ (see equation \eqref{eq:t-trans}).
Thus, the reconstruction of $u_2$ can 
{\it in principle} be obtained by inverting the
set of linear equations \eqref{eq:t-trans} \cite{LAS15}. 

\begin{theorem}
The coefficients of $u_2\in\hilbert_2$ in \eqref{eq:u2form}
satisfy the following system of linear equations
\begin{equation}
\mbox{\boldmath$\Xi$}\mathbf{d} = \mathbf{q},
~~~\text{where}~~~
\mbox{\boldmath$\Xi$} =
[ (\mbox{\boldmath$\Xi$}^1)^T , ... , (\mbox{\boldmath$\Xi$}^M)^T ]^T
~~~\text{and}~~~
\mathbf{q} = [ (\mathbf{q}^1)^T ,..., (\mathbf{q}^M)^T ]^T 
\label{eq:decode_eq}
\end{equation}
with
$
\left[\mathbf{q}^i\right]_k = q^i_k,
\left[\mathbf{d}\right]_{l_{t_1}l_{t_2}} = d_{l_{t_1}l_{t_2}} 
$
and
\[
\left[\mbox{\boldmath$\Xi$}^i\right]_{k; l_{t_1} l_{t_2}} =
\int_{t^i_k}^{t^i_{k+1}}\!\!\!e_{l_{t_1}+l_{t_2}}(t) dt
\int_{\mathbb{D}^2} h^i_2(s_1;s_2)
e_{-l_{t_1}}(s_1)
e_{-l_{t_2}}(s_2) ds_1ds_2.
\]
\label{thrm:volterra-tdm}
\end{theorem}
The above result can be obtained by plugging \eqref{eq:u2form} into \eqref{eq:t-trans}. We refer readers to Theorem 1 in
\cite{LAS15} for a detailed proof. 

We formulate the reconstruction of $u_2$ as the following optimization problem:
\begin{equation}
\hat{u}_2(t_1;t_2) = \underset{u_2 \in \hilbert_2}{\operatorname{argmin}} \sum_{i=1}^{M}\sum_{k\in \mathbb{I}^i} ( \mathcal{T}^i_k u_2 - q^i_k)^2.
\label{eq:plain_rec}
\end{equation}

\begin{algorithm}
	The solution to \eqref{eq:plain_rec} is given by
	
	\begin{equation}
		\hat{u_2}(t_1;t_2) =\!\!\! \sum_{l_{t_1}=-L_t}^{L_t}\sum_{l_{t_2}=-L_t}^{L_t} \hat{d}_{l_{t_1} l_{t_2}} e_{l_{t_1}}(t_1) \cdot e_{l_{t_2}}(t_2),
		\label{eq:solution_plain}
	\end{equation}
	where $\hat{\mathbf{d}} = [\hat{d}_{{-L_t},{-L_t}},\cdots,\hat{d}_{{-L_t},{L_t}}, \cdots, \cdots, \hat{d}_{{L_t},{-L_t}}, \cdots, \hat{d}_{{L_t},{L_t}}]^T$ is obtained by
	\begin{equation}
	\hat{\mathbf{d}} = \mbox{\boldmath$\Xi$}^\dag \mathbf{q}
	\end{equation}
	with $^\dag$ denoting the pseudoinverse operator.
\label{al:volterra-tdm}
\end{algorithm}

We note that a necessary condition for perfect recovery is that the total number of spikes exceeds $dim(\hilbert_1)(dim(\hilbert_1)+1)/2+M$ \cite{LAY14a}.
Therefore, the complexity
of the decoding algorithm is on the order of $dim(\hilbert_1)^2$.

Following \cite{LAS15, LAY14a}, the decoding algorithm is called a Volterra Time
Decoding Machine (Volterra TDM).

\subsection{Functional Identification of DSPs of Complex Cells}
\label{sec:volterra_cim}

In this section, we formulate the functional identification of a single complex cell in the neural circuit described in Figure~\ref{fig:cc_square}.
We perform $M$ experimental trials.
In trial $i, i=1,\cdots,M$, we present a controlled stimulus $u^i_1(t)$ to the cell and observe the spike times $(t^i_k)_{k\in\mathbb{I}^i}$. We assume the cell has a DSP of the from $h_2(t_1;t_2) = g^1_1(t_1)g^1_1(t_2) + g^2_1(t_1)g^2_1(t_2)$ and an integrate and fire BSG with integration constant, bias and threshold denoted by $\kappa, b \mbox{ and } \delta$, respectively.  The objective
is to functionally identify $h_2$ from the knowledge of $u_1^i$ and the observed spikes $(t^i_k)_{k\in\mathbb{I}^i}$, $i=1,\cdots,M$.
This is a standard practice in neurophysiology for inferring the functional form
of a component of a sensory system \cite{HW1962}.

\begin{definition}
Let $h_p \in \mathbb{L}^1(\mathbb{D}^p), p=1,2$, where $\mathbb{L}^1$ denotes the space of Lebesgue integrable functions. The operator $\mathcal{P}_1: \mathbb{L}_1(\mathbb{D}) \rightarrow \hilbert_1$ given by
\begin{equation}
(\mathcal{P}_1 h_1)(t) = \int_{\mathbb{D}} h_1(t') K_1(t; t') dt'
\end{equation}
is called the projection operator from $\mathbb{L}^1(\mathbb{D})$ to $\hilbert_1$.
Similarly, the operator $\mathcal{P}_2: \mathbb{L}_1(\mathbb{D}^2) \rightarrow \hilbert_2$ given by
\begin{equation}
(\mathcal{P}_2h_2)(t_1;t_2) = \int_{\mathbb{D}^2} h_2(t'_1; t'_2) K_2(t_1,t_2; t'_1,t'_2) dt'_1dt'_2
\end{equation}
is called the projection operator from $\mathbb{L}^1(\mathbb{D}^2)$ to $\hilbert_2$.
\end{definition}

Note, that for $u_1^i \in \hilbert_1, \mathcal{P}_1 u_1^i = u_1^i$. Moreover, with $u_2^i(t_1,t_2) = u_1^i(t_1)u_1^i(t_2),  \mathcal{P}_2 u_2^i = u_2^i$.

\begin{lemma}
With $M$ trials of stimuli $u^i_2(t_1;t_2) = u^i_1(t_1)u^i_1(t_2), i=1,\cdots,M$, presented to a complex cell having DSP $h_2(t_1,t_2)$, we have
	\begin{equation}
	\mathcal{L}^i_k (\mathcal{P}_2 h_2) = q_k^i, k\in\mathbb{I}^i, i = 1,\cdots, M,
	\label{eq:id_encode}
	\end{equation}
where
\begin{equation}
\mathcal{L}^i_k (\mathcal{P}_2 h_2) = \int_{t^i_k}^{t^i_{k+1}} \int_{\mathbb{D}^2}  u^i_2(t-s_1; t-s_2) (\mathcal{P} h_2) (t-s_1;t-s_2) ds_1 ds_2 dt,
\label{eq:id_functional_form}
\end{equation}
and
\begin{equation}
q_k^i =  \kappa^i\delta^i - b^i(t^i_{k+1}-t^i_{k}).
\label{eq:id_q}
\end{equation}
\label{lem:dual}
\end{lemma}
\proof 
With \eqref{eq:id_q} the t-transform for the $i^{th}$ stimulus is given by
\[
\int_{t^i_k}^{t^i_{k+1}} \int_{\mathbb{D}^2} h_2 (t-s_1;t-s_2) u^i_2(s_1; s_2) ds_1 ds_2 dt = q_k^i.
\]
Since $\mathcal{P}_2 u_2^i = u_2^i$, we have
\begin{flalign*}
& \int_{t^i_k}^{t^i_{k+1}} \int_{\mathbb{D}^2} h_2 (t-s_1;t-s_2) (\mathcal{P}_2 u^i_2)(s_1; s_2) ds_1 ds_2 dt = q_k^i ~~~\text{or}&\\
&
\int_{t^i_k}^{t^i_{k+1}} \int_{\mathbb{D}^2}\int_{\mathbb{D}^2} h_2 (t-s_1;t-s_2)K_2(s_1,s_2; s'_1,s'_2)  u^i_2(s'_1; s'_2) ds'_1ds'_2 ds_1 ds_2 dt = q_k^i ~~~\text{or}&\\
&
\int_{t^i_k}^{t^i_{k+1}} \int_{\mathbb{D}^2}\int_{\mathbb{D}^2} h_2 (t-s_1;t-s_2)K_2(t-s_1,t-s_2; t-s'_1,t-s'_2)ds_1 ds_2  u^i_2(s'_1; s'_2) ds'_1ds'_2  dt = q_k^i ~~~\text{or}& \\
&
\int_{t^i_k}^{t^i_{k+1}} \int_{\mathbb{D}^2} (\mathcal{P}_2h_2) (t-s_1;t-s_2) u^i_2(s_1; s_2) ds_1ds_2  dt = q_k^i . &\\
\end{flalign*}
Finally, with \eqref{eq:id_functional_form},
we obtain
\begin{equation}
	\mathcal{L}^i_k (\mathcal{P}_2 h_2) = q_k^i, k\in\mathbb{I}^i, i = 1,\cdots, M.
	\label{eq:iden_t}
\end{equation}
\qed

\begin{figure}[htbp] 
\centering
\subfloat[]{ \label{fig:dual_dec} \includegraphics[width=0.44\textwidth]{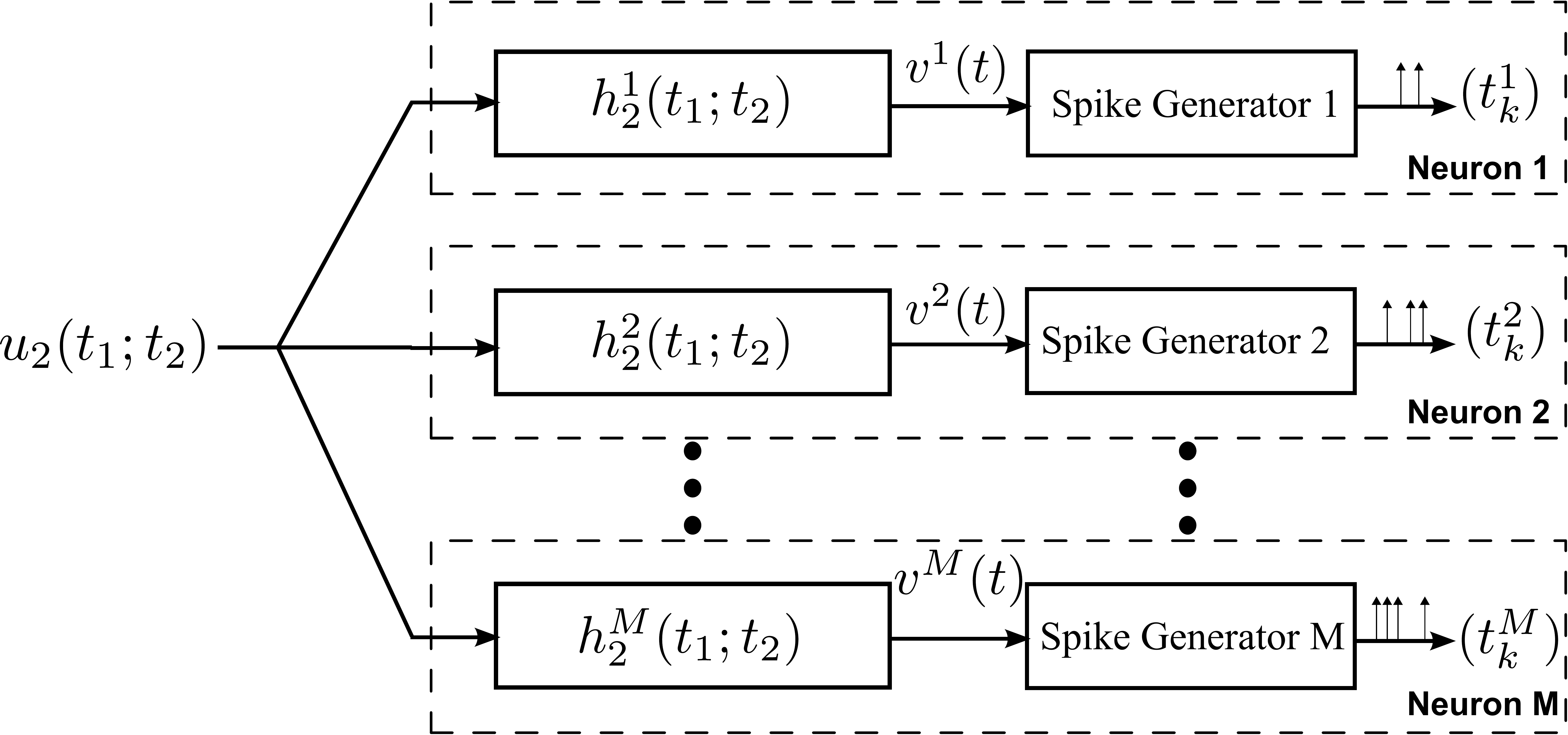}}~%
\subfloat[]{ \label{fig:dual_id} \includegraphics[width=0.465\textwidth]{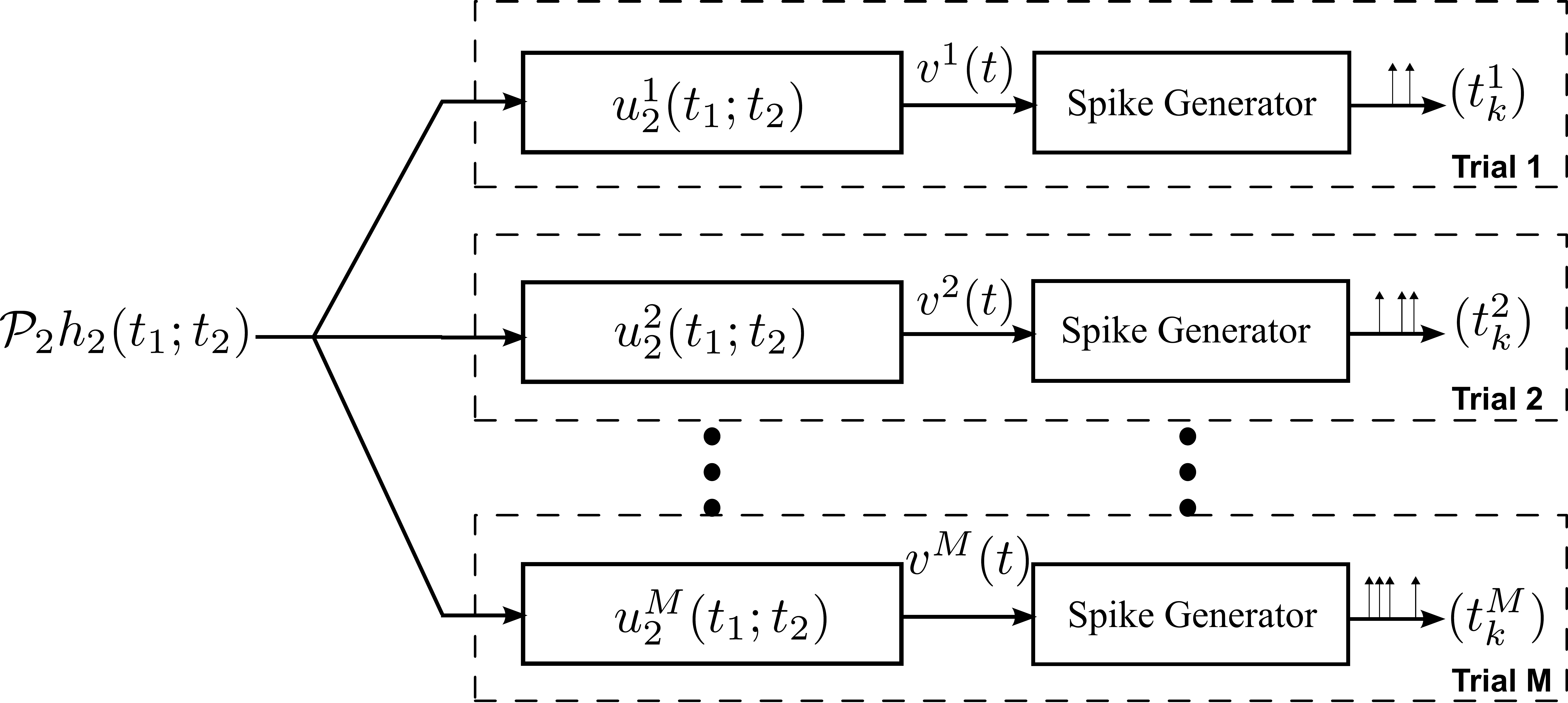}}
   \caption{Duality between decoding and identification. (a) The stimulus $u_1(t)$
   is encoded with a population of complex cells. (b) The projection of the second-order Volterra DSP of an arbitrary neuron on the input space generates the same spike trains if the impulse responses of the DSPs are the same as the input stimuli in repeated trials.}                                                                                                          
   \label{fig:dual_1}
\end{figure}

\begin{remark}
The similarity between equations~\eqref{eq:t-trans} and~\eqref{eq:iden_t} suggests that the identification of a complex cell DSP by presenting multiple stimuli is dual to decoding a stimulus encoded by a population of complex cells. This duality is schematically shown
in Figure~\ref{fig:dual_1}. 
\end{remark}

\begin{theorem}
Let $\mathcal{P}_2h_2 \in\hilbert_2$ be of the form
\begin{equation}
\mathcal{P}_2h_2(t_1;t_2) = \!\!\! \sum_{l_{t_1}=-L_t}^{L_t}\sum_{l_{t_2}=-L_t}^{L_t} h_{l_{t_1} l_{t_2}} e_{l_{t_1}}(t_1) \cdot e_{l_{t_2}}(t_2) .
\label{eq:h2form}
\end{equation}
Then,
$
\left[\mathbf{h}\right]_{l_{t_1}l_{t_2}} = h_{l_{t_1}l_{t_2}}$ with 
$l_{t_1} = -L_t,\cdots,L_t, l_{t_2} = -L_t,\cdots,L_t$, 
satisfies the following system of linear equations
\begin{equation}
\mbox{\boldmath$\Theta$}\mathbf{h} = \mathbf{q},
\end{equation}
where
$\mbox{\boldmath$\Theta$} =
[ (\mbox{\boldmath$\Theta$}^1)^T , ... , (\mbox{\boldmath$\Theta$}^M)^T ]^T$ 
and
$\mathbf{q} = [ (\mathbf{q}^1)^T ,..., (\mathbf{q}^M)^T ]^T $
with
$\left[\mathbf{q}^i\right]_k = q^i_k,$
and
\begin{equation}
\left[\mbox{\boldmath$\Theta$}^i\right]_{k; l_{t_1} l_{t_2}} =
\int_{t^i_k}^{t^i_{k+1}}\!\!\!e_{l_{t_1}+l_{t_2}}(t) dt
\int_{\mathbb{D}^2} u^i_2(s_1;s_2)
e_{-l_{t_1}}(s_1)
e_{-l_{t_2}}(s_2) ds_1ds_2.
\end{equation}
\label{thrm:volterra-cim}
\end{theorem}

Thus, to identify $\mathcal{P}_2 h_2$, we can follow the same methodology as in Algorithm~\ref{al:volterra-tdm}, and formulate the functional identification of $\mathcal{P}_2 h_2$ as
\begin{equation}
\widehat{\mathcal{P}_2 h_2} = \underset{\mathcal{P}_2 h_2 \in \hilbert_2}{\operatorname{argmin}} \sum_{i=1}^{M}\sum_{k\in\mathbb{I}^i}\left( \mathcal{L}^i_k(\mathcal{P}_2 h_2) - q^i_k \right)^2.
\label{eq:plain_id}
\end{equation}

\begin{algorithm}
	The solution to \eqref{eq:plain_id} is given by
	\begin{equation}
		\widehat{\mathcal{P}_2h_2}(t_1;t_2) =\!\!\! \sum_{l_{t_1}=-L_t}^{L_t}\sum_{l_{t_2}=-L_t}^{L_t} \hat{h}_{l_{t_1} l_{t_2}} e_{l_{t_1}}(t_1) \cdot e_{l_{t_2}}(t_2),
		\label{eq:solution_plain_id}
	\end{equation}
	where $\hat{\mathbf{h}} = [\hat{h}_{{-L_t},{-L_t}},\cdots,\hat{h}_{{-L_t},{L_t}}, \cdots, \cdots, \hat{h}_{{L_t},{-L_t}}, \cdots, \hat{h}_{{L_t},{L_t}}]^T$ is obtained by
	\begin{equation}
	\hat{\mathbf{h}} = \mbox{\boldmath$\Theta$}^\dag \mathbf{q}.
	\end{equation}
\label{al:volterra-cim}
\end{algorithm}

The methodology described in Algorithm~\ref{al:volterra-cim} to identify the nonlinear DSP is called the Volterra Channel Identification Machine (Volterra CIM)
\cite{LAS15, LAY14a}.

\begin{remark}
	Formulating the decoding and identification problems in the tensor product space $\hilbert_2$ allows the identification of nonlinear
processing by solving a set of linear equations. However, the increased dimensionality necessitates the use of $\mathcal{O}\left(dim\left(\hilbert_1\right)^2\right)$ measurements.
\end{remark}

\section{Low-Rank Decoding and Functional Identification}
\label{sec:rankmin}

As shown in Section~\ref{sec:VTDM}, a reconstruction of the signal $u_2$ is in principle
possible by solving a set of linear equations.
However, the complexity of the algorithm is prohibitive.
We show in this section that an efficient decoding algorithm
can be constructed that exploits the structure of encoding circuits with complex cells.
Based on the duality between decoding and functional identification,
functional identification algorithms that exploit the structure of the DSP
of complex cells are presented.
These algorithms largely reduce the complexity
of decoding of temporal stimuli encoded by an ensemble of complex cells
and that of functional identification of their DSPs.

\subsection{Low-Rank Decoding of Stimuli}
\label{sec:low_rank}

\subsubsection{Exploiting the Structure of Complex Cell Encoding }

In Theorem~\ref{thrm:volterra-tdm}, we introduced a vector notation for the coefficients of $u_2$
\begin{equation}
\mathbf{d} = [d_{{-L_t},{-L_t}},\cdots,d_{{-L_t},{L_t}}, \cdots, \cdots, d_{{L_t},{-L_t}}, \cdots, d_{{L_t},{L_t}}]^T.
\end{equation}
We introduce here the matrix notation of the coefficients for $u_2 \in \hilbert_2$,
\begin{equation}
\mathbf{D} =
\left[
\begin{array}{ccc}
d_{{-L_t},{L_t}} & \ldots & d_{{-L_t},{-L_t}} \\
\vdots & \ddots & \vdots \\
d_{{L_t},{L_t}} & \ldots & d_{{L_t},{-L_t}}
\end{array}
\right].
\label{eq:D_matrix_form}
\end{equation}
We notice the following: i)
since $u_2$ is assumed to be real, $\overline{d_{l_{t_1},l_{t_2}}} = d_{-l_{t_1},-l_{t_2}}$,
and ii) since $u_2(t_1;t_2) = u_1(t_1)u_1(t_2) = u_1(t_2)u_1(t_1) = u_2(t_2;t_1)$,
we have $d_{l_{t_1},l_{t_2}} = d_{l_{t_2},l_{t_1}} $.
These properties imply that $\mathbf{D}$ is 
a Hermitian matrix.
Moreover, we
note that $u_2$ in \eqref{eq:t-trans} is the ``outer"
product of the stimuli $u_1$, \textit{i.e.},
\begin{equation}
\mathbf{D} = \mathbf{cc}^H,
\label{eq:outer_product}
\end{equation}
where
\begin{equation}
\mathbf{c}= \left[ c_{-L_t}, \cdots, c_{ L_t} \right]^T
\label{eq:c_vector_form}
\end{equation}
are the coefficients of the basis functions of $u_1$.
Therefore, $\mathbf{D}$ is a rank-1 Hermitian positive semidefinite matrix. This property will be exploited in stimulus decoding (reconstruction).

\begin{theorem}
Encoding the stimulus $u_1\in\hilbert_1$
with the neural circuit with complex cells given in \eqref{eq:iaf} into the spike train sequence
$(t_k^i), k\in\mathbb{I}^i$, $i=1,2,...,M$, satisfies the set of equations
\begin{equation}
\mbox{\bf Tr}( \mbox{\boldmath$\Phi$}^i_k \mathbf{D}) = q^i_k, k \in \mathbb{I}^i, i = 1,\cdots, M,
\label{eq:trace-t-complex}
\end{equation}
where $\mbox{\bf Tr}(\cdot)$ is the trace operator,
$\mathbf{D}$ is the rank-$1$ positive semidefinite Hermitian matrix
 $\mathbf{D} = \mathbf{c}\mathbf{c}^H$, $q^i_k=\kappa^i\delta^i - b^i (t^i_{k+1}-t^i_{k})$
and $(\mbox{\boldmath$\Phi$}^i_k), k \in \mathbb{I}^i, i = 1,\cdots, M$, are Hermitian matrices with entries in the
$\left(l_{t_2}+L_t+1\right)$-th row and $\left(l_{t_1}+L_t+1\right)$-th column given by
\begin{equation}
[\mbox{\boldmath$\Phi$}^i_k]_{l_{t_2}, l_{t_1}} = \int_{t^i_k}^{t^i_{k+1}} e_{l_{t_1}-l_{t_2}}(t)dt \int_{\mathbb{D}^2} h^i_2(s_1;s_2) e_{-l_{t_1}}(s_1)e_{l_{t_2}}(s_2) ds_1ds_2 .
\label{eq:phi}
\end{equation}
\label{th:trace-trans}
\end{theorem} 
\proof 
Plugging in the general form of $u_2$ in \eqref{eq:u2form} into \eqref{eq:operatorT}, 
the left hand side of \eqref{eq:t-trans} amounts to
\[
\sum_{l_{t_1}=-L_t}^{L_t}\sum_{l_{t_2}=-L_t}^{L_t}
d_{l_{t_1},- l_{t_2}} 
\int_{t^i_k}^{t^i_{k+1}}e_{l_{t_1}-l_{t_2}}(t) dt
\int_{\mathbb{D}^2} h^i_2(s_1;s_2)e_{-l_{t_1}}(s_1)e_{l_{t_2}}(s_2) ds_1ds_2.
\]
It is easy to verify that the above expression can be written as
\begin{equation}
\sum_{l_{t_1}=-L_t}^{L_t}\sum_{l_{t_2}=-L_t}^{L_t} d_{l_{t_1},- l_{t_2}}
[\mbox{\boldmath$\Phi$}^i_k]_{l_{t_2}, l_{t_1}} =
\mbox{\bf Tr}( \mbox{\boldmath$\Phi$}^i_k\mathbf{D} ).
\label{eq:trace}
\end{equation}
Finally, we note that since $h^i_2, i=1,\cdots,M$, are assumed to be real valued, $(\mbox{\boldmath$\Phi$}^i_k), k\in\mathbb{I}^i, i=1,\cdots,M$, are Hermitian. \qed

\begin{remark}
We note that equation \eqref{eq:trace-t-complex} in Theorem~\ref{th:trace-trans}
and equation \eqref{eq:decode_eq} in Theorem~\ref{thrm:volterra-tdm} 
are the same.
These equations represent the t-transform of a complex cell in (rank-1) matrix and vector form,
respectively.
The (rank-1) matrix representation is made possible by the equality $u_2(t_1;t_2) = u_1(t_1)u_1(t_2)$.
\end{remark}

\subsubsection{Reconstruction Algorithms}
\label{sec:t_dec}

Solving the systems of equations \eqref{eq:trace-t-complex} and \eqref{eq:decode_eq} requires at least $dim(\hilbert_1)(dim(\hilbert_1)+1)/2+M$ measurements.
Consequently, practical solutions become quickly intractable.
Fortunately, the encoded stimulus is of the form $u_2(t_1;t_2) = u_1(t_1)u_2(t_2)$.
This guarantees that $\mathbf{D}$ is a rank-1 matrix and thus the reconstructed
stimulus belongs to a small subset of $\hilbert_2$.
Therefore, we can cast the problem of reconstructing temporal stimuli encoded by neural circuits with
complex cells as a feasibility problem, that is, find all positive semidefinite Hermitian matrices that satisfy \eqref{eq:trace-t-complex} and have rank 1.
As we shall demonstrate, the latter condition can be satisfied with substantially fewer measurements.

Recently, there is an increasing interest in low-rank optimizations
such as matrix factorization, matrix completion and rank minimization,
both from a theoretical and from a practical standpoint \cite{CZP09, RFP10, FHB2004}.
For example, rank minimization has recently been applied to 
phase retrieval problems \cite{CSV2011}.

Our objective here is to find rank-1, positive-semidefinite matrices that satisfy 
the t-transform \eqref{eq:trace-t-complex}.
Since there always exists at least one rank-1 solution,
this is equivalent to the following optimization problem \cite{CES13}
\begin{equation}
\begin{array}{cc}
\mbox{minimize} & \mbox{\bf Rank}{(\mathbf{D})} \\
\mbox{s.t.} & \mbox{\bf Tr}( \mbox{\boldmath$\Phi$}^i_k \mathbf{D}) = q^i_k, k \in \mathbb{I}^i, i = 1,\cdots, M,\\
& \mathbf{D} \succcurlyeq 0
\end{array}
\label{eq:vtem_min_rank}
\end{equation}

The rank minimization problem in \eqref{eq:vtem_min_rank} is NP-hard.
A well known heuristic is to relax the problem \eqref{eq:vtem_min_rank}
to a trace minimization problem \cite{FHB2004}.
That is, instead of solving \eqref{eq:vtem_min_rank}, we reconstruct $u_2$ using Algorithm~\ref{al:rankmin}.
\begin{algorithm}
The reconstruction of $u_2$ from the spike times generated by the neural circuit with complex cells is given by
\begin{equation}
		\hat{u_2}(t_1;t_2) =\!\!\! \sum_{l_{t_1}=-L_t}^{L_t}\sum_{l_{t_2}=-L_t}^{L_t} \hat{d}_{l_{t_1} l_{t_2}} e_{l_{t_1}}(t_1) \cdot e_{l_{t_2}}(t_2),
		\label{eq:recon_D}
	\end{equation}
where 
\begin{equation}
\hat{\mathbf{D}} =
\left[
\begin{array}{ccc}
\hat{d}_{{-L_t},{L_t}} & \ldots & \hat{d}_{{-L_t},{-L_t}} \\
\vdots & \ddots & \vdots \\
\hat{d}_{{L_t},{L_t}} & \ldots & \hat{d}_{{L_t},{-L_t}}
\end{array}
\right].
\end{equation}
is the solution to the semidefinite programming (SDP) problem
\begin{equation}
\begin{array}{cc}
\mbox{minimize} & \mbox{\bf Tr}{(\mathbf{D})} \\
\mbox{s.t.} & \mbox{\bf Tr}( \mbox{\boldmath$\Phi$}^i_k \mathbf{D}) = q^i_k, k\in\mathbb{I}^i, i = 1,\cdots, M\\
& \mathbf{D} \succcurlyeq 0
\end{array},
\label{eq:vtem_min_tr}
\end{equation}
\label{al:rankmin}
\end{algorithm}

When the matrices $(\mbox{\boldmath$\Phi$}^i_k)$,
$ k\in\mathbb{I}^i, i = 1,\cdots, M$, satisfy the rank restricted isometry property \cite{RFP10}, the trace norm relaxation converges to the true solution of \eqref{eq:vtem_min_rank} provided that the number of measurements is of the order $\mathcal{O}\Big(dim(\mathcal{H}_1)log\big(dim\left(\mathcal{H}_1\right)\big)\Big)$ \cite{RFP10}. 
These results suggest that stimuli encoded by complex cells can be decoded with a significantly lower number of measurements than that required by Algorithm~\ref{al:volterra-tdm}.
To investigate this further, we applied the above algorithm to decode a large number of stimuli encoded by complex cells while varying the number of measurements (spikes) used by the decoding algorithm. The results show that the number of spikes required to faithfully represent a stimulus by a neural circuits consisting of complex cells is quasilinearly rather than quadratically proportional to the dimension of the stimulus space. These results are presented in the subsequent sections.

The matrix of weights $\hat{\mathbf{D}}$ obtained from the above algorithm
can be further decomposed to extract the signal $u_1$ (up to a sign) as follows.\\
(i) Perform the eigen-decomposition of
$\hat{\mathbf{D}}$.
Denote the largest eigenvalue by
$\lambda$ and the corresponding eigenvector by
$\mathbf{v}$.
If \eqref{eq:vtem_min_tr} 
does not exactly return a rank-$1$ matrix, choose the
largest eigenvalue and disregard the rest.
Let $\mathbf{w} = \sqrt{\lambda}\mathbf{v}$.\\
(ii)
The reconstructed stimulus $\hat{u}_1$ is given by (up to a sign)
\[
\hat{u}_1(t) = \sum_{l_t=-L_t}^{L_t} \hat{c}_{l_t} e_{l_t}(t),
\]
where
\begin{equation}
\hat{\mathbf{c}} = \left\{ \begin{array}{cc} 
 \mathbf{w}\cdot \frac{| [\mathbf{w}]_{L_t+1} |}{[\mathbf{w}]_{L_t+1}}, &  \mbox{ if } [\mathbf{w}]_{L_t+1} \neq 0 \\ 
 \mathbf{w} , & \mbox{otherwise} 
\end{array} \right.
\end{equation}
with
$\hat{\mathbf{c}}= \left[ \hat{c}_{ -L_t}, \cdots, \hat{c}_{L_t}\right]^T$,
and
$[\mathbf{w}]_{L_t+1}$ is the $(L_t+1)^{\text{th}}$ entry of $\mathbf{w}$, which corresponds to the coefficient $\hat{c}_0$.

If $\hat{\mathbf{D}}$ is rank 1, step (i) decomposes $\hat{\mathbf{D}}$ as an ``outer" product of a vector and itself (see \eqref{eq:outer_product}).
The resulting vector $\mathbf{w}$ differs from the actual coefficient vector of the stimulus $u_1$ by up to a complex-valued scaling factor.
This factor is corrected in step (ii). Since $u_1$ is assumed to be real-valued, the ``DC" component must be real-valued.
Therefore, we rotate $\mathbf{w}$ to remove any imaginary part. In practice, this also ensures $\hat{c}_{-l_t} = \overline{\hat{c}_{l_t}}$.

\begin{remark}
Note that we can reconstruct $u_1(t)$ up to a sign, since $\mathbf{D} = \mathbf{c}\mathbf{c}^H$ and $\mathbf{D} = (-\mathbf{c})(-\mathbf{c}^H)$ are equally possible.\end{remark}
\begin{remark} 
Note that \eqref{eq:vtem_min_rank} can be alternatively solved by replacing the objective with the log-det heuristic \cite{FHB2004},
that is
\vspace{-0.075in}
\begin{equation}
\begin{array}{cc}
\mbox{minimize} & log~det(\mathbf{D}+\lambda \mathbf{I}) \\
\mbox{s.t.} & \mbox{\bf Tr}( \mbox{\boldmath$\Phi$}^i_k \mathbf{D}) = q^i_k,k\in\mathbb{I}^i, i = 1,\cdots, M,\\
& \mathbf{D} \succcurlyeq 0
\end{array}
\label{eq:vtem_min_logdet}
\end{equation}
where $\lambda > 0$ is a small regularization constant.
This optimization may further reduce the rank of $\hat{\mathbf{D}}$ when Algorithm~\ref{al:rankmin} fails to
progress to an exact rank-1 solution. \cite{FHB2004}.
\end{remark}

While the SDP in \eqref{eq:vtem_min_tr} provides an elegant way for relaxing the rank minimization problem,
it is limited in practice by the need of large amounts of computer memory for numerical calculations. 
The optimization problem \eqref{eq:vtem_min_rank} can also be solved using an alternating minimization scheme \cite{JNS2013} as outlined in Algorithm~\ref{al:altmin} below.
The alternating minimization approach is more tractable when the dimension of the
space is very large.
Algorithm~\ref{al:altmin} uses an initialization step (step $1$ below) that provides an initial iterate whose distance from $\mathbf{D}$ is bounded.
It then alternately solves for the left and right singular vector of the rank-1 matrix $\mathbf{D}$ while keeping the other one fixed (step $2$ below).
The resulting subproblems admit a straightforward least squares solution, that can be much more efficiently solved than the SDP in Algorithm~\ref{al:rankmin}. Moreover, the algorithm is amenable to parallel computation using General Purpose Graphics Processing Units (GPGPUs).
The latter property makes it even more attractive when the dimension of the stimulus space is large.

\begin{algorithm}
\begin{enumerate}
\item Initialize $\hat{\mathbf{c}}_1$ and $\hat{\mathbf{c}}_2$ to top left and right singular vector respectively of $\sum_{i=1}^M \sum_{k\in\mathbb{I}^i} q_k^i \mathbf{\Phi_k^i}$ normalized to $\sqrt{\frac{1}{\sigma}\sum_{i=1}^M\sum_{k\in\mathbb{I}^i} (q_k^i)^2}$, where $\sigma$ is the top singular value of $\sum_{i=1}^M \sum_{k\in\mathbb{I}^i} q_k^i \mathbf{\Phi_k^i}$.
\item Solve alternately the following two minimization problems 
	 \begin{enumerate}
		\item solve for $\hat{\mathbf{c}}_1$ by fixing $\hat{\mathbf{c}}_2$
		\begin{equation}
		\hat{\mathbf{c}}_1 = \operatornamewithlimits{min}_{\mathbf{c}_1}  \sum_{i=1}^M \sum_{k\in\mathbb{I}^i} (\mbox{\bf Tr}( \mbox{\boldmath$\Phi$}^i_k \mathbf{c}_1 \hat{\mathbf{c}}_2^H) - q^i_k)^2
		\end{equation}
		\item solve for $\hat{\mathbf{c}}_2$ by fixing $\hat{\mathbf{c}}_1$
		\begin{equation}
		\hat{\mathbf{c}}_2 = \operatornamewithlimits{min}_{\mathbf{c}_2}  \sum_{i=1}^M \sum_{k\in\mathbb{I}^i} (\mbox{\bf Tr}( \mbox{\boldmath$\Phi$}^i_k \hat{\mathbf{c}}_1 \mathbf{c}_2^H) - q^i_k)^2
		\end{equation}
	\end{enumerate}
	until $\sum_{i=1}^M \sum_{\in\mathbb{I}^i} (\mbox{\bf Tr}( \mbox{\boldmath$\Phi$}^i_k \hat{\mathbf{c}}_1 \hat{\mathbf{c}}_2^H ) - q^i_k)^2 \leq \epsilon$, where $\epsilon>0$ is the error tolerance level.
\item compute $\hat{\mathbf{D}} = \hat{\mathbf{c}}_1 \hat{\mathbf{c}}_2^H$.
\end{enumerate}
\label{al:altmin}
\end{algorithm}

$\hat{\mathbf{D}}$ approximates the coefficients of $u_2 \in \hilbert_2$ as in \eqref{eq:recon_D}
We can reconstruct $u_1$, using the (appropriately scaled) top eigenvector of $\frac{1}{2} (\hat{\mathbf{D}} + \hat{\mathbf{D}}^H)$. This can be obtained directly from $\hat{\mathbf{c}}_1$ and $\hat{\mathbf{c}}_2$ as follows. Let
\begin{equation}
 k = \frac{\hat{\mathbf{c}}_1^H \hat{\mathbf{c}}_2-\hat{\mathbf{c}}_2^H \hat{\mathbf{c}}_1 + \sqrt{\left(\hat{\mathbf{c}}_1^H \hat{\mathbf{c}}_2-\hat{\mathbf{c}}_2^H \hat{\mathbf{c}}_1\right)^2 + 4\hat{\mathbf{c}}_1^H \hat{\mathbf{c}}_1\hat{\mathbf{c}}_2^H\hat{\mathbf{c}}_2}}{2\hat{\mathbf{c}}_2^H\hat{\mathbf{c}}_2},
\end{equation}
and
\begin{equation}
	\mathbf{w} = \sqrt{\frac{1}{2} \hat{\mathbf{c}}_2^H \hat{\mathbf{c}}_1 + k \hat{\mathbf{c}}_2^H\hat{\mathbf{c}}_2 } \frac{\hat{\mathbf{c}}_1 + k\hat{\mathbf{c}}_2}{\| \hat{\mathbf{c}}_1 + k\hat{\mathbf{c}}_2  \|},
\end{equation}
the reconstructed stimulus $\hat{u}_1$ is given by (up to a sign)
\[
\hat{u}_1(t) = \sum_{l_t=-L_t}^{L_t} \hat{c}_{l_t} e_{l_t}(t),
\]
where
\begin{equation}
\hat{\mathbf{c}} = \left\{ \begin{array}{cc} 
\mathbf{w}\cdot \frac{| [\mathbf{w}]_{L_t+1} |}{[\mathbf{w}]_{L_t+1}} , & \mbox{    if    } [\mathbf{w}]_{L_t+1} \neq 0,\\
\mathbf{w}, &  \mbox{   otherwise  } 
\end{array} \right.
\end{equation}
with $\hat{\mathbf{c}} = \left[ \hat{c}_{ -L_t}, \cdots, \hat{c}_{L_t}\right]^T$.

We point out that we made the decoding manageable by exploiting the structure of $u_2$.
Therefore, there is no constraint on the exact form $h^i_2(t_1;t_2)$ can take, and
the decoding algorithms can be applied to neural circuits with neurons whose DSPs take
the form of any second-order Volterra kernel.

\subsubsection{Example - Decoding of Temporal Stimuli Encoded with a Population of Complex Cells}
\label{sec:ex_dec}
Here, the neural circuit we consider consists of $19$ complex cells.
The DSPs of the complex cells are of the form
\begin{equation}
h_2^i( t_1;  t_2) = g^{i1}_1 ( t_1) g^{i1}_1( t_2) + g^{i2}_1( t_1) g^{i2}_1( t_2) ,
\end{equation}
where $g^{i1}_1(t)$ and  $g^{i2}_1(t)$  are quadrature pairs of temporal Gabor filters and $i=1,\cdots,19$.
The Gabor filters are constructed from dilations and translations of the mother wavelets, where the mother functions
can expressed as
\begin{equation}
g^1_1(t) = \operatorname{exp}\left(- \left(\frac{t^2}{0.001} \right) \right) \operatorname{cos}\left( 40\pi t \right),
\end{equation}
and
\begin{equation}
g^2_1(t) = \operatorname{exp}\left(- \left(\frac{t^2}{0.001} \right) \right) \operatorname{sin}\left( 40\pi t \right) .
\end{equation}
The BSG of the complex cells are point IAF neurons with bias $b^i = 2$ and integration constant $\kappa^i = 1$, for $i = 1,\cdots,M$.
These two parameters are kept the same for all stimuli. Different threshold values are chosen for the IAF neurons in order
to vary the total number of spikes, which can be used to evaluate how many measurements are required for perfectly
reconstructing the input stimuli.

The domain of the input space $\hilbert_1$ is $\mathbb{D} = [0,1]$ (sec) and 
$L_t = 20, \Omega_t = 20\cdot 2\pi$ (rad/sec). Thus, we have $dim(\hilbert_1) = 41$.
 The stimuli were generated by randomly choosing their basis coefficients from an i.i.d. Gaussian distribution.

We tested the encoding and subsequent decoding of $6,570$ stimuli. The total number of spikes produced for each stimulus ranged from 20 to 220. Reconstructions of the stimuli were performed using Algorithm~\ref{al:rankmin}, and
the SDPs were solved using SDPT3 \cite{TTT2003}.

\begin{figure}[htbp] 
   \centering
   \subfloat[]{\label{fig:dec_all_temp}\includegraphics[width=0.545\textwidth]{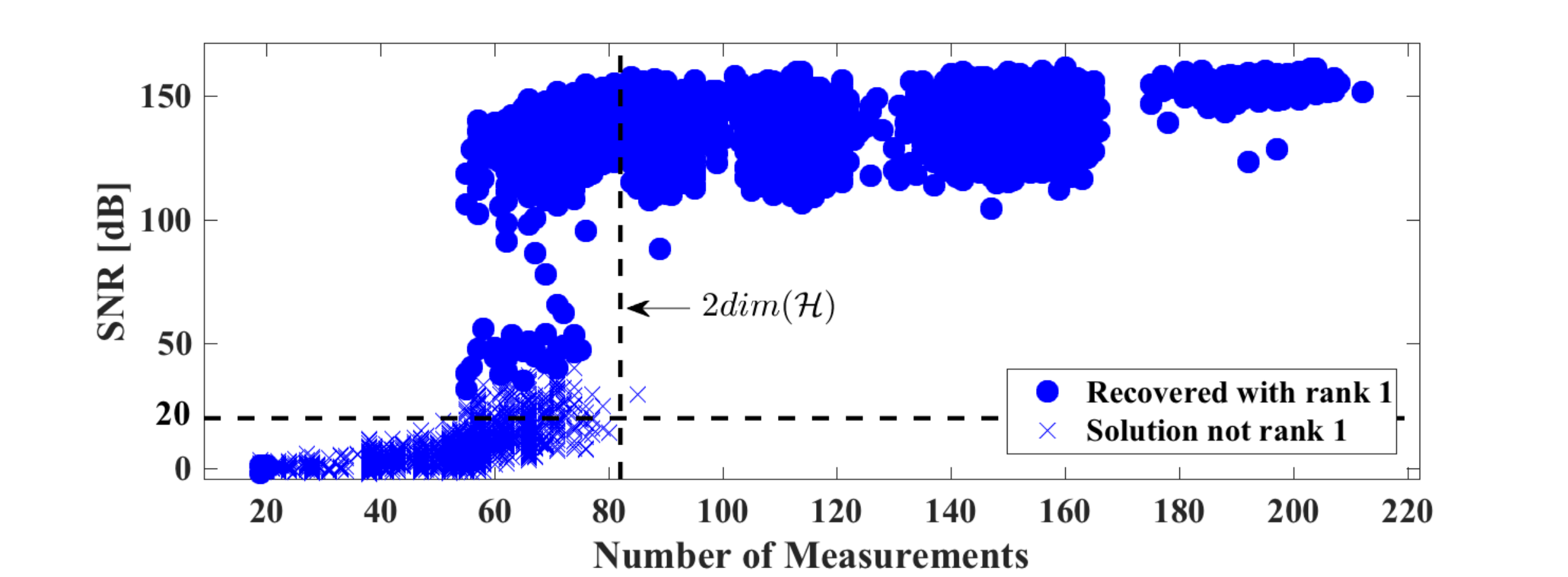}}
   \subfloat[]{\label{fig:dec_rate_temp}\includegraphics[width=0.385\textwidth]{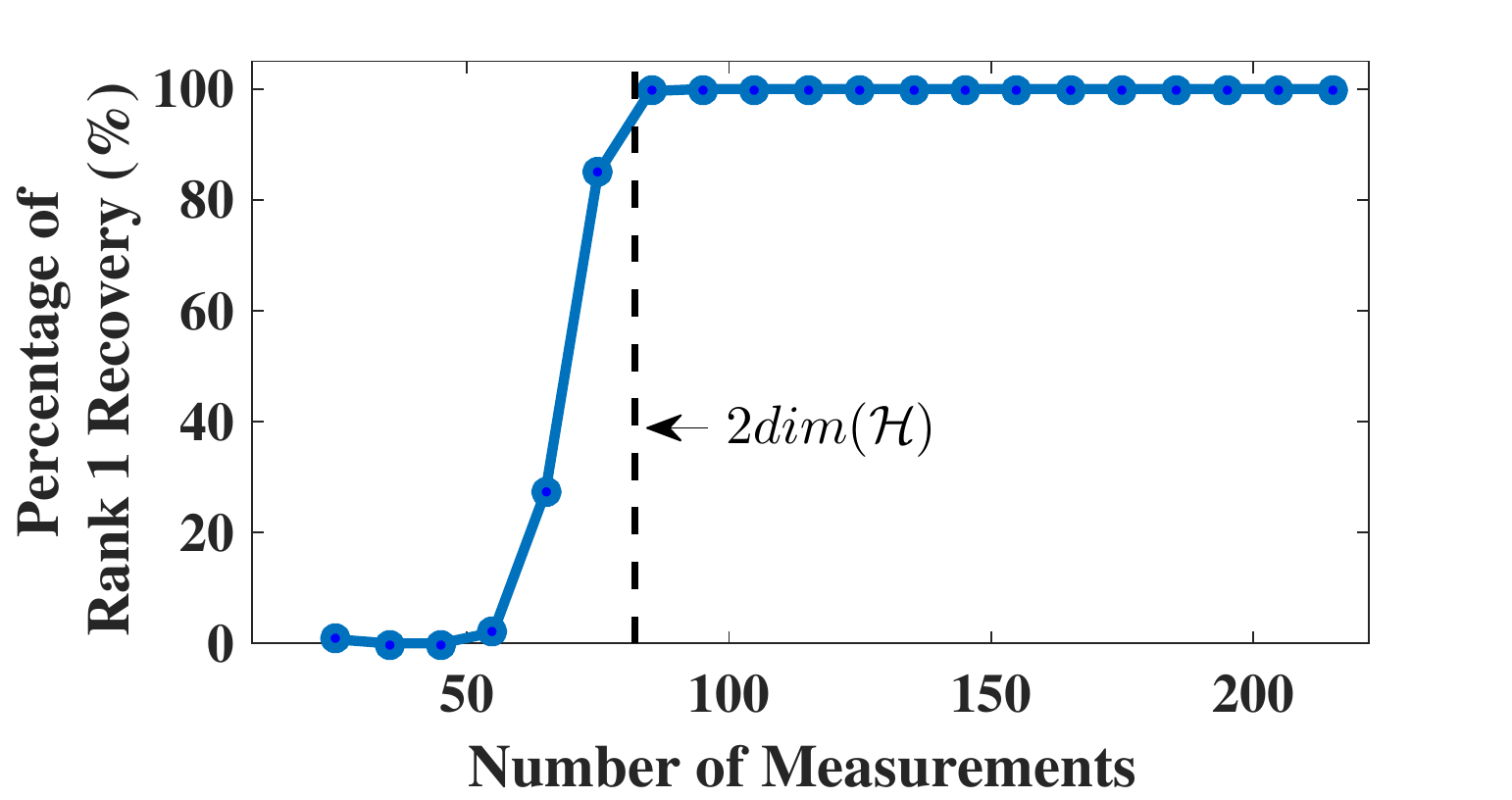}}
   \caption{Example of low-rank decoding. (a) Effect of number of measurements (spikes) on reconstruction quality. (b) Percentage of rank 1 reconstructions.}
   \label{fig:dec_bound}
\end{figure}

We show the SNR of all reconstructions in the scatter plot of Figure~\ref{fig:dec_all_temp}. Here solid dots represent
exact rank 1 solutions (largest eigenvalue is at least 100 times larger than the sum of the rest of the
eigenvalues), and crosses indicate that the trace minimization found a higher rank solution that has a
smaller trace. The percentage of exact rank 1 solutions is shown in Figure~\ref{fig:dec_rate_temp}. 
A relatively sharp transition from very low probability of recovery to very high rate of perfect reconstruction
can be seen, similar to phase transition phenomena in other sparse recovery algorithms \cite{DMM09}.
It can also be seen that the
number of measurements that are needed for perfect recovery is substantially lower than the $861$ spikes required by decoding based on Theorem~\ref{al:volterra-tdm}.

\subsection{Low-Rank Functional Identification of Complex Cells}
\label{sec:identification}

\subsubsection{Duality Between Low-Rank Functional Identification and Decoding}

As discussed in Section~\ref{sec:volterra_cim}, the complexity of  identification using Algorithm~\ref{al:volterra-cim}
can be prohibitively high.
Often, a very large number
of stimulus presentation trials are required to fully identify the
DSP of biological neurons. 
To mitigate this, we consider exploiting the structure of the DSP of complex cells as motivated by the tractability of the low rank decoding algorithm when the structure of the stimuli is explored.
 
We consider a single complex cell whose
DSP is of the form
\begin{equation}
h_2(t_1;t_2) = \sum_{n=1}^N g^{n}_1(t_1)g^{n}_1(t_2),
\label{eq:low_rank_dsp}
\end{equation}
where $g^n_1(t), n=1,\cdots,N$, are impulse responses of linear filters,
and $N \ll dim(\hilbert_1)$.
We note that a complex cell described in Figure~\ref{fig:cc_square} is a special case of \eqref{eq:low_rank_dsp} with $N=2$.
A natural question here is that, by assuming such a structure,
 whether the functional identification of 
 complex cell DSPs is tractable.

\begin{remark}
It is well known that a second-order Volterra kernel has infinite equivalent forms but has a unique symmetric form \cite{Rugh1981}.
\end{remark}

We have shown that the low-rank structure of $u_2$ leads to a reduction of complexity
in the reconstruction of temporal stimuli encoded by an ensemble of complex cells.
We also described the duality between decoding and functional identification. If we can show that the functional identification formalism
for complex cell DSP is the dual to decoding of low-rank stimuli, it is straightforward
to provide tractable algorithms for identifying $h_2(t_1;t_2)$ of the form \eqref{eq:low_rank_dsp}.

Since $\mathcal{P}_1g^n_1(t) \in \hilbert_1, n=1,\cdots,N$,  there is a set of coefficients $(g^n_{l_t}), l_t = -L_t,...,L_t$ and
$n=1,2,...,N$, such that
\begin{equation}
\mathcal{P}_1g^n_1(t) = \sum_{l_t=-L_t}^{L_t} g^n_{l_t}e_{l_t}(t).
\end{equation}
In what follows wte denote coefficients in vector form as
\begin{equation}
\mathbf{g}^n = \left[g^n_{-L_t}, \cdots, g^n_{L_t}  \right]^T .
\label{eq:pg_vector_form}
\end{equation}
Similarly, we denote the coefficients of $\mathcal{P}_1h_2(t_1;t_2)$
in \eqref{eq:h2form}
in matrix form as
\begin{equation}
\mathbf{H} =
\left[
\begin{array}{ccc}
h_{{-L_t},{L_t}} & \ldots & h_{{-L_t},{-L_t}} \\
\vdots & \ddots & \vdots \\
h_{{L_t},{L_t}} & \ldots & h_{{L_t},{-L_t}}
\end{array}
\right].
\label{eq:ph_matrix_form}
\end{equation}
Then
\begin{equation}
\mathbf{H} = \sum_{n=1}^N \mathbf{g}^n (\mathbf{g}^n)^H
\end{equation}
and thus $\mathbf{H}$
is a Hermitian positive semidefinite matrix with rank at most $N$.

\begin{theorem}
By presenting $M$ trials of stimuli $u^i_2(t_1;t_2) = u^i_1(t_1)u^i_1(t_2), i=1,\cdots,M$
to a complex cell, its coefficients satisfy the set of equations
\begin{equation}
\mbox{\bf Tr}( \mbox{\boldmath$\Psi$}^i_k \mathbf{H}) = q^i_k, k\in\mathbb{I}^i, i = 1,\cdots, M,
\label{eq:trace-t-id}
\end{equation}
where $n_i+1, i=1,\cdots,M$, is the number of spikes generate by the complex cell in trial $i$,
$\mathbf{H}$ is a Hermitian positive semidefinite matrix with $\operatorname{rank}(\mathbf{H}) \leq N$,
given by
$
\mathbf{H} = \sum_{n=1}^N \mathbf{g}^n(\mathbf{g}^n)^H,
$
with
$\mathbf{g}^n= \left[ g^n_{-L_t}, \cdots, g^n_{L_t} \right]^T,$
$(\mbox{\boldmath$\Psi$}^i_k), k\in\mathbb{I}^i, i = 1,\cdots, M$, are Hermitian matrices with entry at
$\left(l_{t_2}+L_t+1\right)$-th row and $\left(l_{t_1}+L_t+1\right)$-th column given by
\begin{equation}
[\mbox{\boldmath$\Psi$}^i_k]_{l_{t_2}; l_{t_1}} =
\int_{t^i_k}^{t^i_{k+1}} e_{l_{t_1}-l_{t_2}}(t)dt \int_{\mathbb{D}^2} u_2^i(s_1;s_2)e_{ -l_{t_1}}( s_1)e_{ l_{t_2}}( s_2) ds_1ds_2.
\label{eq:psi}
\end{equation}
\label{th:dual-t}
\end{theorem} 
\proof 
From Lemma~\ref{lem:dual}, we have
\begin{equation}
\mathcal{L}^i_k (\mathcal{P}_2h_2) = q^i_k, k\in \mathbb{I}^i, i=1,\cdots,M,
\end{equation}
where
\begin{equation}
\mathcal{L}^i_k (\mathcal{P}_2h_2) = \int_{t^i_k}^{t^i_{k+1}} \int_{\mathbb{D}^2} u^i_2(t-s_1;t-s_2) (\mathcal{P}_2h_2)(s_1;s_2) ds_1 ds_2 dt.
\end{equation}

\eqref{eq:trace-t-id} can be obtained  following the steps of the proof of Theorem~\ref{th:trace-trans}. \qed

\begin{figure}[htbp] 
\centering
\subfloat[]{ \label{fig:dual_dec} \includegraphics[width=0.43\textwidth]{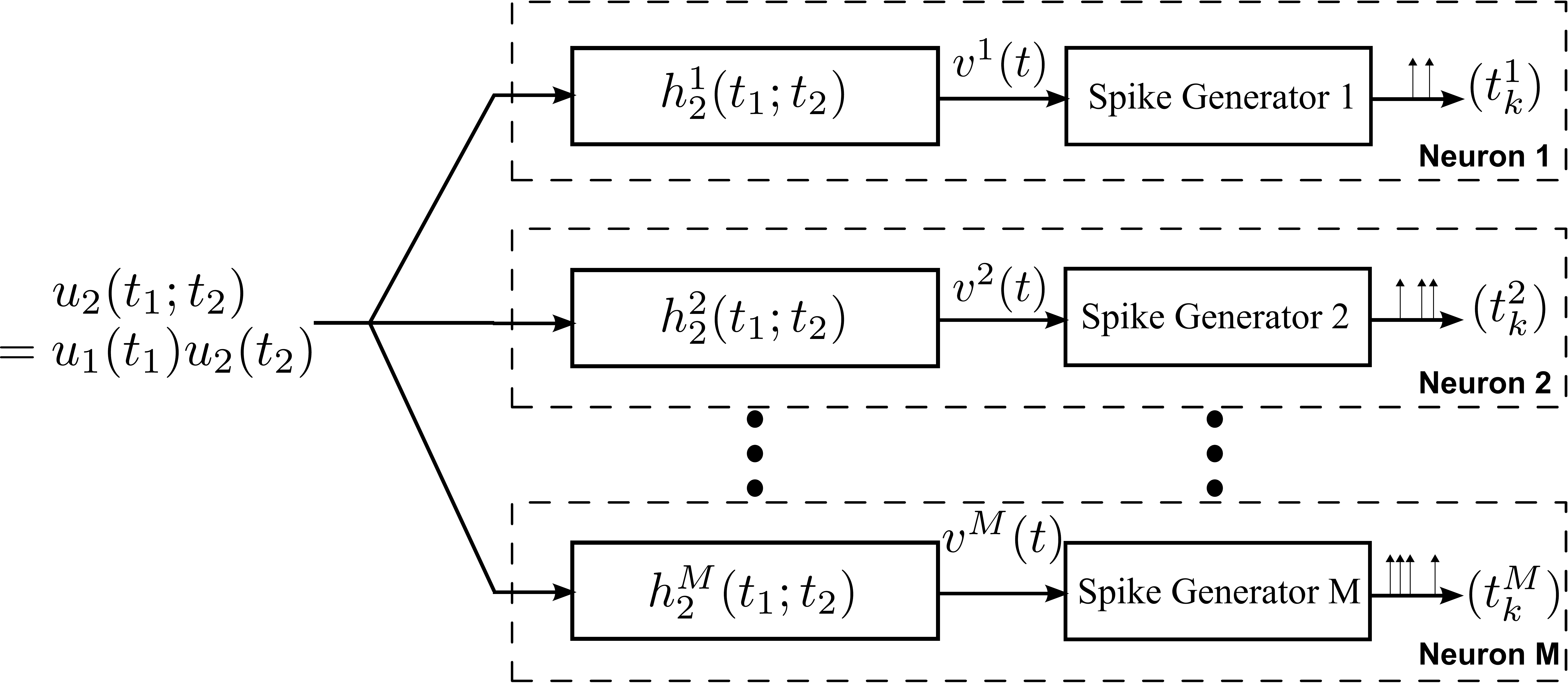}}~%
\subfloat[]{ \label{fig:dual_id} \includegraphics[width=0.495\textwidth]{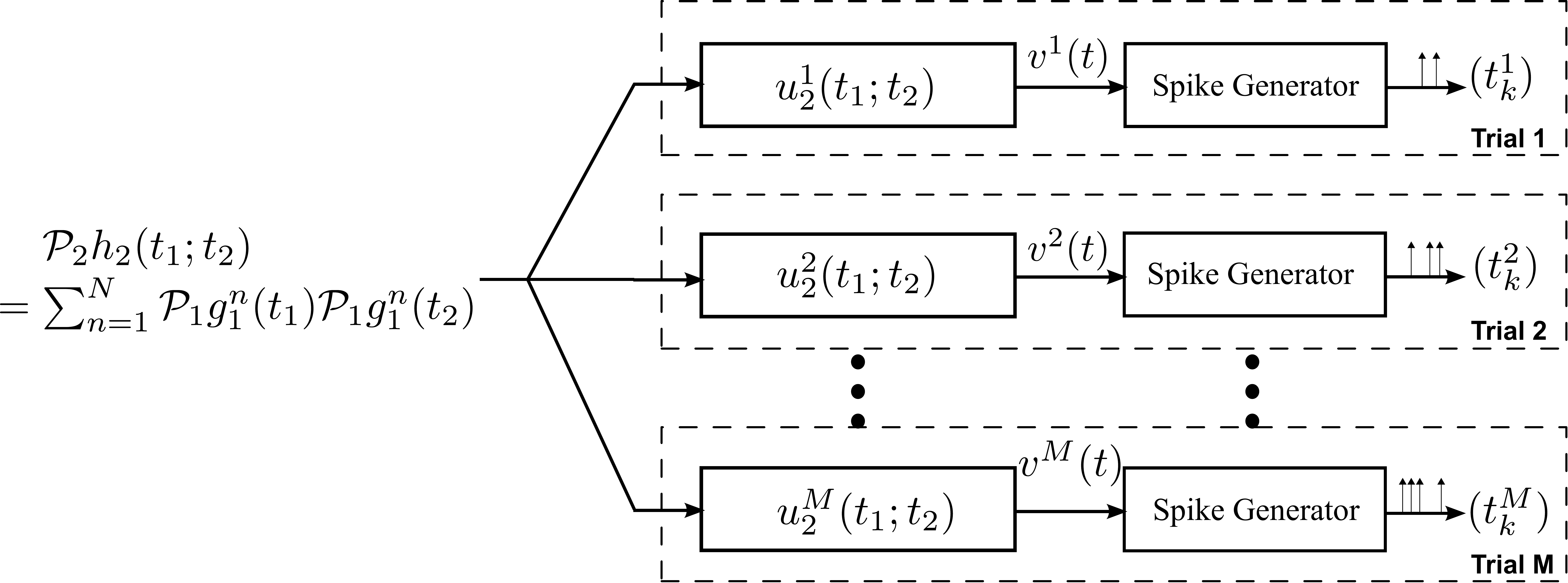}}
   \caption{Duality between low-rank decoding of a stimulus encoded by a  population of complex cells
   and low-rank functional identification of complex cells. (a) The low-rank decoding algorithm assumes that the
   encoded stimulus can be written as $u_2(t_1;t_2)=u_1(t_1)u_1(t_2)$.
   (b) Functional identification of a complex cell assumes that the structure of the DSP is low rank, \textit{i.e.},
   $\mathcal{P}_2h_2(t_1;t_2) = \sum_{n=1}^N \mathcal{P}_1g^n_1(t_1)\mathcal{P}_1g^n_1(t_2)$.}
   \label{fig:dual}
\end{figure}

\begin{remark}
As in Section~\ref{sec:identification}, we note that the similarity in \eqref{eq:trace-t-id} and \eqref{eq:trace-t-complex} indicates the duality between low-rank functional identification of complex cells and low-rank decoding of stimuli encoded by a population of complex cells. The duality is illustrated in Figure~\ref{fig:dual}.
\end{remark}

\subsubsection{Functional Identification Algorithms}
\label{sec:t_id}

To functionally identify the complex cell  DSP,
we again employ a rank minimization problem
\begin{equation}
\begin{array}{cc}
\mbox{minimize} & \mbox{\bf Rank}{(\mathbf{H})} \\
\mbox{s.t.} & \mbox{\bf Tr}( \mbox{\boldmath$\Psi$}^i_k \mathbf{H}) = q^i_k, k\in\mathbb{I}^i, i = 1,\cdots, M\\
& \mathbf{H} \succcurlyeq 0
\end{array},
\label{eq:cim_min_rank}
\end{equation}
Algorithm~\ref{al:rankmin} provides a solution to the above rank minimization problem.
However, in this case, the optimal solution shall have
rank $N$. We relax the problem to a trace minimization problem and consider the following
algorithm for low-rank functional identification of complex cells.

\begin{algorithm}
The functional identification of complex cell DSP from the spike times generated by the neuron in $M$ stimulus trials is given by
\begin{equation}
		\widehat{\mathcal{P}_2h_2}(t_1;t_2) =\!\!\! \sum_{l_{t_1}=-L_t}^{L_t}\sum_{l_{t_2}=-L_t}^{L_t} \hat{h}_{l_{t_1} l_{t_2}} e_{l_{t_1}}(t_1) \cdot e_{l_{t_2}}(t_2),
	\end{equation}
where
\begin{equation}
\hat{\mathbf{H}} =
\left[
\begin{array}{ccc}
\hat{h}_{{-L_t},{L_t}} & \ldots & \hat{h}_{{-L_t},{-L_t}} \\
\vdots & \ddots & \vdots \\
\hat{h}_{{L_t},{L_t}} & \ldots & \hat{h}_{{L_t},{-L_t}}
\end{array}
\right].
\end{equation}
is the solution to the SDP problem
\begin{equation}
\begin{array}{cc}
\mbox{minimize} & \mbox{\bf Tr}{(\mathbf{H})} \\
\mbox{s.t.} & \mbox{\bf Tr}( \mbox{\boldmath$\Psi$}^i_k \mathbf{H}) = q^i_k, k\in\mathbb{I}^i, i = 1,\cdots, M\\
& \mathbf{H} \succcurlyeq 0
\end{array},
\end{equation}
\label{al:rankmin_id}
\end{algorithm}

Based on the results for decoding using Algorithm~\ref{al:rankmin} and provided that $h_2$ is of the form \eqref{eq:low_rank_dsp}, we intuitively inferred that the number of measurements for the perfect identification of $\mathcal{P}_2h_2$  is much smaller than $\mathcal{O}\big(dim(\mathcal{H}_1)^2\big)$ .
We demonstrate that this is the case for a large number of identification examples in the subsequent sections.

This suggests that\ even if the dimension of the input space becomes large, the functional identification
of the DSP of complex cells is still tractable. This result has critical implication for performing neurobiological experiments
to functionally identify complex cells. First, it suggests that a much smaller number of stimulus trials is needed for perfect
identification. Second, the total number of spikes/measurements that needs to be recorded can be significantly reduced.
Both means the duration of experiment can be shortened.

\begin{remark}
Note that only the projection of the DSP
$h_2$ onto the
space of input stimuli can be identified.
\end{remark}

\begin{remark}
We can use the largest $N$ eigenvalues
and their respective eigenvectors of $\hat{\mathbf{H}}$ to obtain the
projection of individual linear filter components $\widehat{\mathcal{P}_1g^n_1} ,n=1,\cdots,N$.
However, these components may not directly correspond to $\mathcal{P}_1g^n_1 ,n=1,\cdots,N$, in that
the original projections may not be ``orthogonal", whereas the eigenvalue decomposition imposes orthogonality. 
\end{remark}

As in Algorithm~\ref{al:altmin} when applied for solving the decoding problem, the rank minimization problem above can be solved using alternating minimization, as described in Algorithm~\ref{al:5} below. Here, we solve for the top $N$ left and right singular vectors of $\mathbf{H}$ alternately, where $N$ is the rank of the second order Volterra DSP.
We note that the initialization step is akin to running an algorithm very similar to the spike-triggered covariance (STC) algorithm widely used in neuroscience \cite{paninski2003convergence,schwartz2002characterizing,PS2006,SPR2006,park2011bayesian}. 
The subsequent steps then  improve upon this initial estimate. 

\begin{algorithm}
\begin{enumerate}
\item Initialize $\hat{\mathbf{H}}_1$ and $\hat{\mathbf{H}}_2$ to top $N$ left and right singular vectors, respectively, of  $\sum_{i=1}^M \sum_{k=1}^{n_i} q_k^i \mathbf{\Psi_k^i}$ with the $n^{th}$ singular vector normalized to
$\frac{1}{N}\sqrt{\frac{1}{\sigma_n}\sum_{i=1}^M\sum_{k=1}^{n_i} (q_k^i)^2}$, where $\sigma_n$ is the top $n^{th}$ singular value of $\sum_{i=1}^M \sum_{k=1}^{n_i} q_k^i \mathbf{\Psi_k^i}$.
\item Solve the following two minimization problems 
	 \begin{enumerate}
		\item solve for $\hat{\mathbf{H}}_1$ by fixing $\hat{\mathbf{H}}_2$
		\begin{equation}
		\hat{\mathbf{H}}_1 = \operatornamewithlimits{min}_{\mathbf{H}_1 \in \mathbb{C}^{dim(
		\mathcal{H}_1) \times N}}  \sum_{i=1}^M \sum_{k\in\mathbb{I}^i} (\mbox{\bf Tr}( \mbox{\boldmath$\Psi$}^i_k \mathbf{H}_1 \hat{\mathbf{H}}_2^H) - q^i_k)^2
		\end{equation}
		\item solve for $\hat{\mathbf{H}}_2$ by fixing $\hat{\mathbf{H}}_1$
		\begin{equation}
		\hat{\mathbf{H}}_2 = \operatornamewithlimits{min}_{\mathbf{H}_2 \in \mathbb{C}^{dim(
		\mathcal{H}_1) \times N}}  \sum_{i=1}^M \sum_{k\in\mathbb{I}^i} (\mbox{\bf Tr}( \mbox{\boldmath$\Psi$}^i_k \hat{\mathbf{H}}_1 \mathbf{H}_2^H) - q^i_k)^2
		\end{equation}
	\end{enumerate}
	until $\sum_{i=1}^M \sum_{\in\mathbb{I}^i} (\mbox{\bf Tr}( \mbox{\boldmath$\Psi$}^i_k \hat{\mathbf{H}}_1 \hat{\mathbf{H}}_2^H ) - q^i_k)^2 \leq \epsilon$, where $\epsilon>0$ is the error tolerance level.
\item compute $\hat{\mathbf{H}} = \frac{1}{2}\left( \hat{\mathbf{H}}_1 \hat{\mathbf{H}}_2^H + \hat{\mathbf{H}}_2 \hat{\mathbf{H}}_1^H \right)$.
\end{enumerate}
\label{al:5}
\end{algorithm}

\subsubsection{Example - Identification of Complex Cell DSPs from Spike Times}
\label{sec:ex_id}

In this example, we consider identifying a single complex cell having the following Volterra DSP
\begin{equation}
h_2(t_1,t_2) = g^1_1(t_1)g^1_1(t_1) + g^2_1(t_1)g^2_1(t_2) ,
\end{equation}
where
\begin{align}
	g^1_1(t) &= 50\;\exp\left(-\frac{(t-0.3)^2}{0.002}\right)\;\cos\left(40\pi t\right), \\
	g^2_1(t) &= 50\;\exp\left(-\frac{(t-0.3)^2}{0.002}\right)\;\sin\left(40\pi t\right).
\end{align}

In repeated trials we presented to the complex cell 1-second long stimuli chosen from the input space.
The domain of the input space $\hilbert^1_1$ is $\mathbb{D} = [0,1]$ (sec) and 
$L_t = 20, \Omega_t = 20\cdot 2\pi$ (rad/sec) and thus, $dim(\hilbert^1_1) = 41$.
The stimuli were generated by independently choosing their basis coefficients from the same Gaussian distribution.
We presented a total of $16,600$ different stimuli  in the repeated trials.
We then randomly selected between 30-80 trial subsets such that the total number of spikes
in each subset was between $60$ and $160$.
We performed the identification process on each subset using Algorithm~\ref{al:rankmin_id}. 
The optimization problem was solved using SDPT3. 

\begin{figure}[h] 
   \centering
  \subfloat[]{\label{fig:id_ex_rate_temp2}\includegraphics[width=0.48\textwidth]{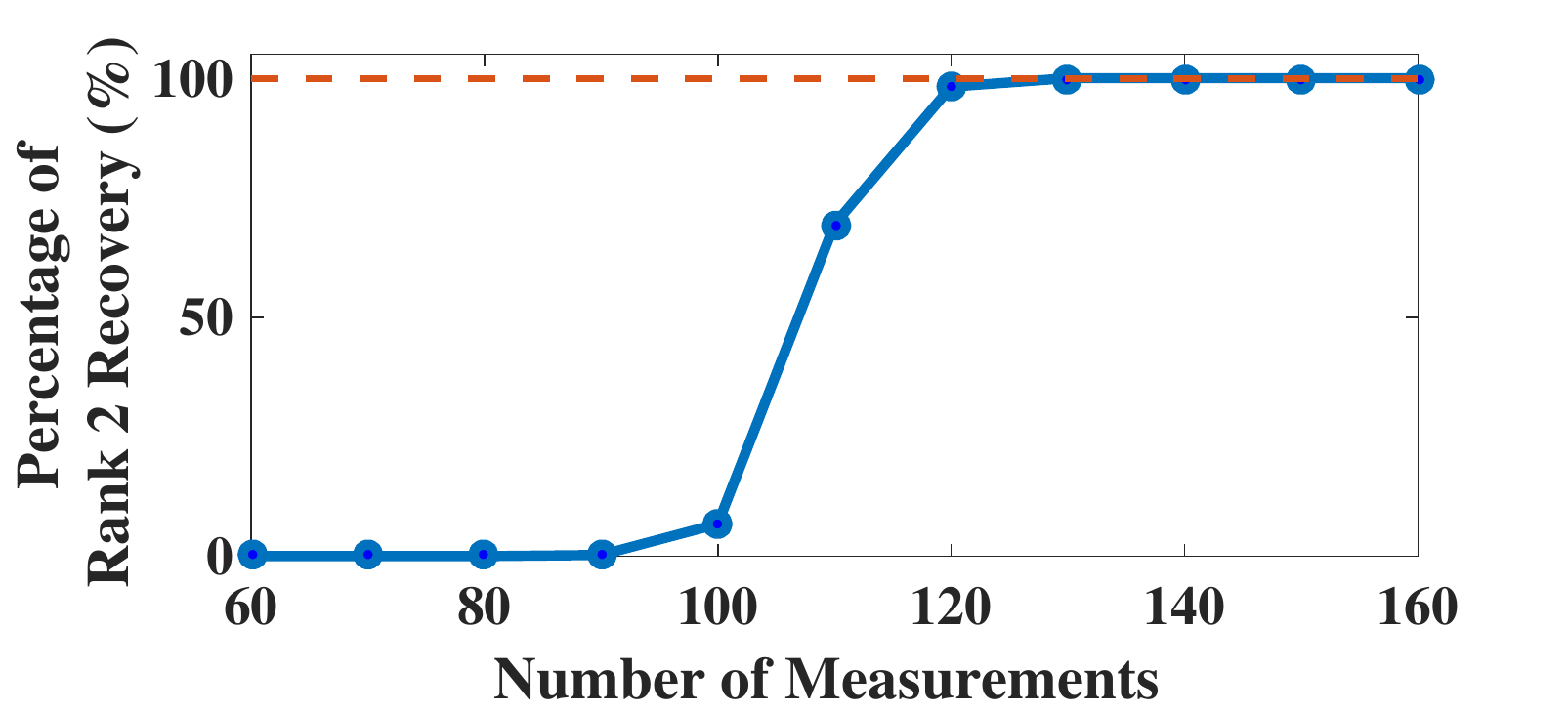}}~~~
  \subfloat[]{\label{fig:id_ex_snr_temp}\includegraphics[width=0.48\textwidth]{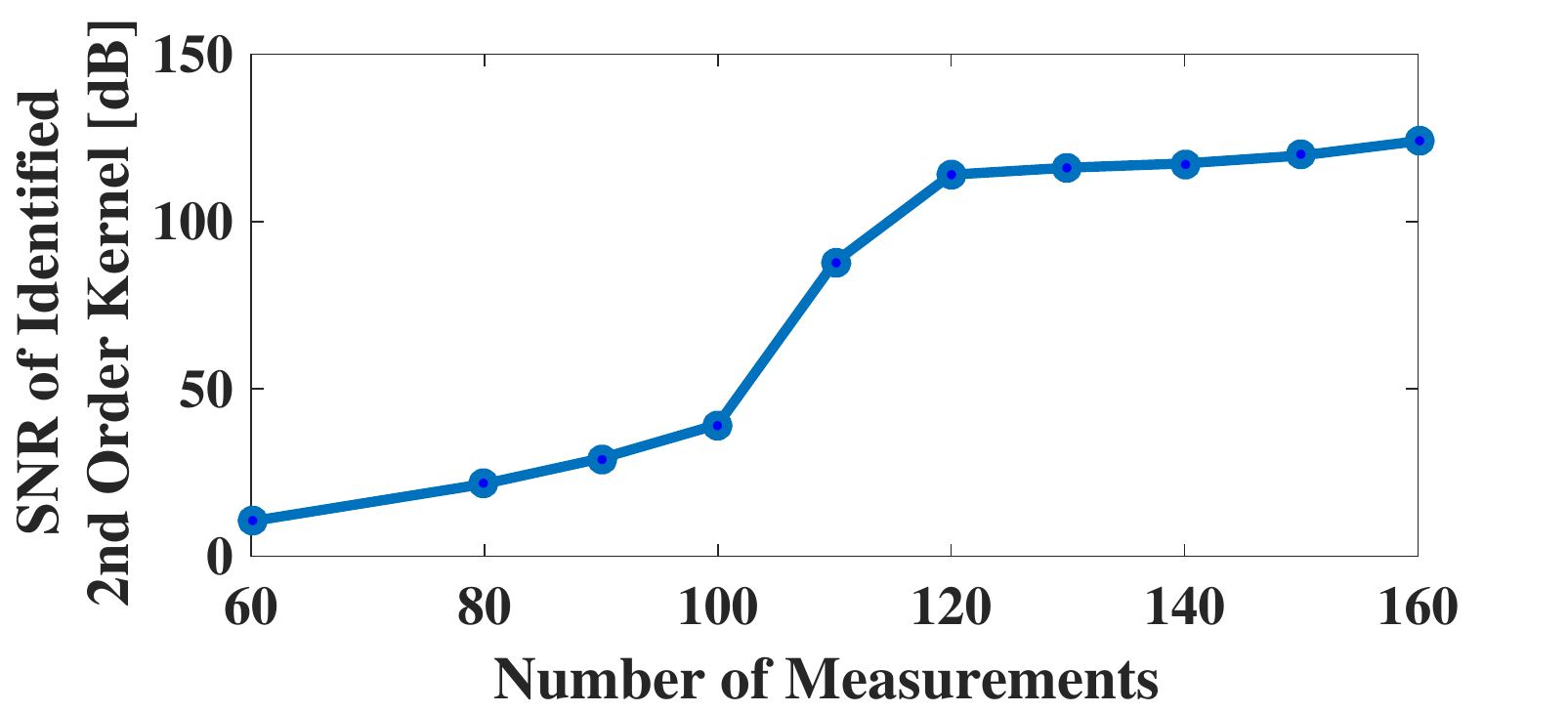}} 
   \caption{Example of low-rank functional identification. (a) Percentage of successful rank-2 recovery in identification. (b) Mean SNR of identified second order DSP kernel.}
\label{fig:id_ex_temp}
\end{figure}

For each instantiation of the identification algorithm, we recorded whether the optimization process resulted in a rank-2 solution and also the SNR of the identified DSP with respect to the original one.
For the purpose of demonstration, we binned these results based on number of spikes used into bins of width $10$.
The percentage of rank-2 solutions is shown in Figure~\ref{fig:id_ex_rate_temp2} as a function of number of measurements. 
The mean SNR is shown in Figure~\ref{fig:id_ex_snr_temp}.

It can be seen from Figure~\ref{fig:id_ex_snr_temp} that the identification algorithm presented here is able to recover the underlying DSP with exceptional accuracy using a reasonable and tractable number of measurements.

\subsection{Evaluation of Functional Identification of a Neural Circuit of Complex Cells by Decoding}

In Section~\ref{sec:low_rank}, we have shown that the sparse decoding algorithm requires much less
number of neurons and measurements (spikes) in the reconstruction of stimuli encoded by
a neural circuit of complex cells. We have also demonstrated in Section~\ref{sec:identification} that
the proposed sparse functional identification algorithm enables the identification of complex cells with
a tractable number of measurements. Together, the two algorithms afford us tractable functional identification
of an entire neural circuit of complex cells that is capable of fully representing stimuli information,
in that i) the size of the neural circuit is tractable, and ii) the requirement for functional identification
is tractable.

In \cite{LSZ2015} and \cite{LAY16}, it was shown that the evaluation of functional identification of an entire
neural circuit can be more intuitively performed in the input space by decoding the stimuli with identified
circuit parameters.
Here, we extend the previous results and apply such evaluation procedure on the sparse decoding and
sparse functional identification algorithms. 
The procedure is described as follows. First, each complex cell is functionally identified using
Algorithm~\ref{al:rankmin_id} or Algorithm~\ref{al:5}. Second, novel stimuli are presented to 
the neural circuit. Third, the spike trains observed are used to reconstruct the encoded novel stimuli
by the sparse decoding algorithm, assuming that the circuit parameters take the identified values.
Finally, SNR of the reconstruction can be obtained. A high SNR indicates a well identified
circuit while a low number implies that the functional identification of the neural circuit is not of good quality.
The latter can be caused by a lack of number of measurements used in functional identification,
or by a lack of complex cells in the neural circuit.

\begin{figure}[h] 
   \centering
	\includegraphics[width=0.8\textwidth]{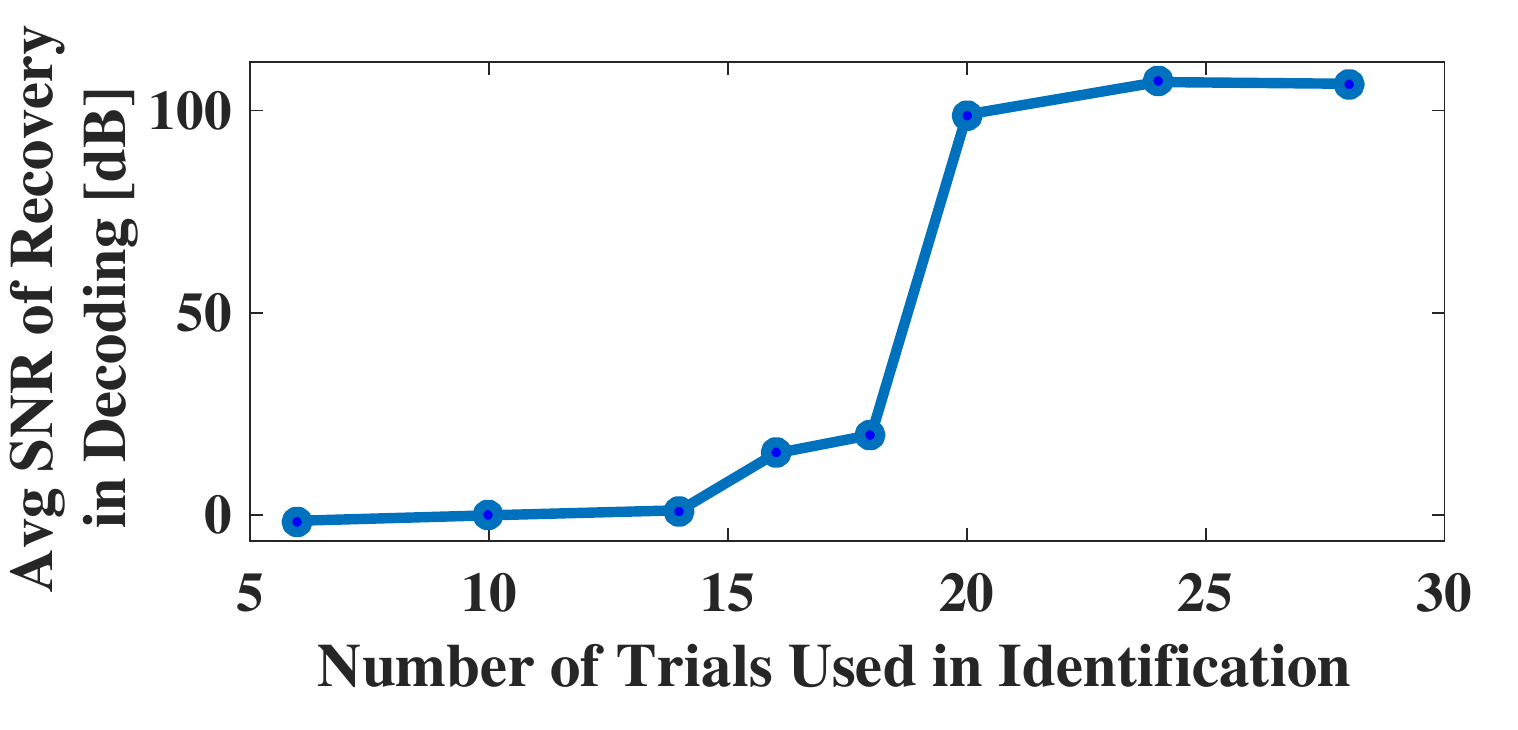}
   \caption{  Evaluating identification quality in the input space by plotting the average SNR of reconstruction of novel stimuli
assumed to be encoded with the identified DSPs.
}
\label{fig:id_temp_decode}
\end{figure}

We performed the functional identification of all $19$ complex cells in the neural circuit given in the example in Section \ref{sec:ex_dec}.
We first identified all complex cells by presenting to the neural circuit $M$ temporal stimuli.
We repeated the  identification of the entire circuit using $8$ different values of $M$.
We then presented to the same circuit (with the original DSPs as in Section \ref{sec:ex_dec}), 100 novel stimuli drawn from the input space and used the spike times
generated by the neural circuit to decode the stimuli.
In the decoding process however, we assumed that the DSPs
of the set of complex cells are as identified, for all $8$ values of $M$.
The mean reconstruction SNR of the $100$ stimuli is shown in Figure~\ref{fig:id_temp_decode}.
As shown, the quality of reconstruction is low until enough trials were used in identification.
When more than 19 trials were performed, perfect reconstruction of the entire neural circuit was achieved.
The dimension of the stimulus space was $41$ and the average number of spikes per neuron used for identification varied from $44$ for $6$ trials to $202$ for $28$ trials.

\section{Low-Rank Decoding and Functional Identification of Complex Cells with Spatio-Temporal Stimuli}
\label{sec:extension}
 
In this section, we extend our results obtained in Section~\ref{sec:rankmin} to neural circuits
with complex cells that encode spatio-temporal stimuli.
We will first introduce the space of spatio-temporal stimuli in Section~\ref{sec:stmodel}.
In Section~\ref{sec:st_decoding}, we formulate the encoding of spatio-temporal stimuli by a population of complex cells.
By extending Theorem \ref{th:trace-trans} to Theorem \ref{th:trace-trans_3}, 
we argue in Section~\ref{sec:st_dec} that decoding of stimuli encoded by a neural circuit of complex cells 
is tractable for spatio-temporal stimuli.
We then employ extension of the Algorithms \ref{al:rankmin} and \ref{al:altmin} developed in Section~\ref{sec:t_dec} 
and demonstrate their effectiveness with a few examples.
In Section~\ref{sec:st_id}, we investigate the duality between functional identification of spatio-temporal DSPs of complex cells
and decoding of stimuli encoded by complex cells with a bank of spatio-temporal DSPs.
Here we extend Theorem \ref{th:dual-t} to the encoding circuits with spatio-temporal complex cells, Theorem \ref{th:dual-st}. We then  apply extensions of the
Algorithms \ref{al:rankmin_id} and \ref{al:5} developed in Section~\ref{sec:t_id} to  the identification of spatio-temporal DSPs of complex cells and
demonstrate their effectiveness.

\subsection{Modeling of Spatio-Temporal Stimuli}
\label{sec:stmodel}

The stimuli $u_1$ defined here have $p$ spatial dimensions and a single temporal dimension,
\textit{i.e.}, $u_1 = u_1(x_1,x_2,\cdots,x_p, t)$. For simplicity of notation,
we use a compact, vector notation and denote the spatial variables as $\mathbf{x} = (x_1,x_2,\cdots,x_p)$.
When $p=2$, $u_1$ is the usual visual stimulus.

\begin{definition}
The space of trigonometric polynomials 
$\hilbert_1^p$ is the Hilbert space of
complex-valued functions
\begin{equation}
u_1(\mathbf{x},t) = \sum_{\mathbf{l_x}} \sum_{l_t} c_{\mathbf{l_x}l_t}e_{\mathbf{l_x}l_t}(\mathbf{x},t),
\label{eq:uform_3}
\end{equation}
where
\[
\mathbf{l_x} \in \{ (l_{x_1},l_{x_2},\cdots,\l_{x_p}) \in \mathbb{Z}^p | -L_{x_1} \leq l_{x_1} \leq L_{x_1}, -L_{x_2} \leq l_{x_2} \leq L_{x_2}, \cdots, -L_{x_p} \leq l_{x_p} \leq L_{x_p} \}, \; 
\]
\[
l_t\in\{k\in \mathbb{Z} | -L_t \leq k \leq L_t \} 
\]
over the domain $\mathbb{D}$, where, by abuse of notation, $\mathbb{D} =  [0,S_{x_1}] \times [0,S_{x_2}] \times \cdots [0,S_{x_p}] \times [0, S_t]  $ and
$S_t = \frac{2\pi L_t}{\Omega_t} \; ,  S_{x_1} = \frac{2\pi L_{x_1}}{\Omega_{x_1}}  \;,  S_{x_2} = \frac{2\pi L_{x_2}}{\Omega_{x_2}}, \cdots, S_{x_p} = \frac{2\pi L_{x_p}}{\Omega_{x_p}}$.
In addition, 
$e_{\mathbf{l_x}l_t}(\mathbf{x},t) = e_{\mathbf{l_x}}(\mathbf{x})e_{l_t}(t)$ where
\[
e_{\mathbf{l_x}}(\mathbf{x}) =  \frac{1}{\sqrt{\prod_{i=1}^p S_i}}\operatorname{exp}\left(j \; \mathbf{\omega_x}^T  \mathbf{x} \right) \quad;\;\; \mathbf{\omega_x} = \left( \frac{l_{x_1}\Omega_{x_1}}{L_{x_1}}, \frac{l_{x_2}\Omega_{x_2}}{L_{x_2}},\cdots, \frac{l_{x_p}\Omega_{x_p}}{L_{x_p}}  \right)
\]
and
\[
e_{l_t}(t) =  \frac{1}{\sqrt{S_t}}\operatorname{exp}\left(\frac{jl_t\Omega_t}{L_t}t \right).
\]
Here $\Omega_t$ denotes the bandwidth,
and $L_t$ the order of the space in the temporal domain while $\Omega_{x_i}$ and $L_{x_i}$ denote the
bandwidth and order of the space in the $i^{th}$ spatial variable.
Stimuli $u_1\in\hilbert_1^p$ are periodic with periods
$S_t , S_{x_1}, \cdots, S_{x_p}$.
\end{definition}
\vspace{0.1in}
$\hilbert_1^p$ is a Reproducing Kernel Hilbert Space (RKHS)
 with
reproducing kernel (RK)
\begin{equation}
K^p_1(\mathbf{x},t;\mathbf{x}',t') = \sum_{\mathbf{l_x}} \sum_{l_t } e_{\mathbf{l_x}l_t}(\mathbf{x}-\mathbf{x}',t-t').
\end{equation}

We denote the temporal dimension of $\hilbert_1^p$ by $dim_t(\hilbert_1^p) = 2L_t+1 $ and the total dimension  by 
$dim(\hilbert_1^p) = (2L_t+1) \prod_{i=1}^p (2L_{x_i}+1) $.

\begin{definition}
The tensor product space $\hilbert_2^p = \hilbert_1^p \otimes \hilbert_1^p$ is a Hilbert space of complex-valued functions
\begin{equation}
u_2(\mathbf{x_1},t_1;\mathbf{x_2},t_2) =\!\!\!
\sum_{\mathbf{l_{x_1}}}\sum_{l_{t_1}}\sum_{\mathbf{l_{x_2}}} \sum_{l_{t_2}} d_{\mathbf{l_{x_1}}  l_{t_1} \mathbf{l_{x_2}} l_{t_2}} \; e_{\mathbf{l_{x_1}}}(\mathbf{x_1}) \; e_{l_{t_1}}(t_1) \; e_{\mathbf{l_{x_2}}}(\mathbf{x_2}) \;
e_{l_{t_2}}(t_2)
\label{eq:u2form_3}
\end{equation}
over the domain $\mathbb{D}^2$.
\end{definition}

$\hilbert_2^p$ is an RKHS with reproducing kernel
\begin{equation}
\begin{split}
K^p_2(\mathbf{x_1},t_1,\mathbf{x_2},t_2 &; \mathbf{x_1}',t'_1,\mathbf{x_2}',t'_2) =\\
=& \sum_{\mathbf{l_{x_1}}} \sum_{l_{t_1}}\sum_{\mathbf{l_{x_2}}}\sum_{l_{t_2}}  e_{\mathbf{l_{x_1}}}(\mathbf{x_1}-\mathbf{x_1}') \;  e_{l_{t_1}}(t_1-t'_1) \; e_{\mathbf{l_{x_2}}} (\mathbf{x_2}-\mathbf{x_2}') \; e_{l_{t_2}}(t_2-t'_2).
\end{split}	
\end{equation}

Note that $dim(\hilbert_2^p) = (dim(\hilbert_1^p))^2$.

\subsection{Encoding of Spatiotemporal Stimuli with a Population of Complex Cells}
\label{sec:st_decoding}

We consider again a neural circuit consisting of a population of
$M$ neurons modeling a population of complex cells as illustrated in
Figure~\ref{fig:complex}.
The input to the neural circuit is spatiotemporal stimulus 
as defined in Section~\ref{sec:stmodel}. 

The input stimulus $u_1(\mathbf{x},t)$ to neuron $i$ is first processed by two spatio-temporal linear filters
whose impulse responses are denoted, by abuse of notation, as $g^{i1}_1(\mathbf{x},t)$ and $g^{i2}_1(\mathbf{x},t)$, respectively.
The output of the linear filters are squared and summed. The sum $v^i(t)$, as the output of the DSP,
 is then
fed into the BSG of neuron $i$.
The BSG encodes the DSP output into 
the spike train $(t^i_k)_{k\in\mathbb{I}^i}$.
Here $\mathbb{I}^i$ is the spike train index set of neuron $i$.

\begin{figure}[htbp] 
   \centering
\subfloat[]{ \label{fig:dual_st_encode}  \includegraphics[width=0.43\textwidth]{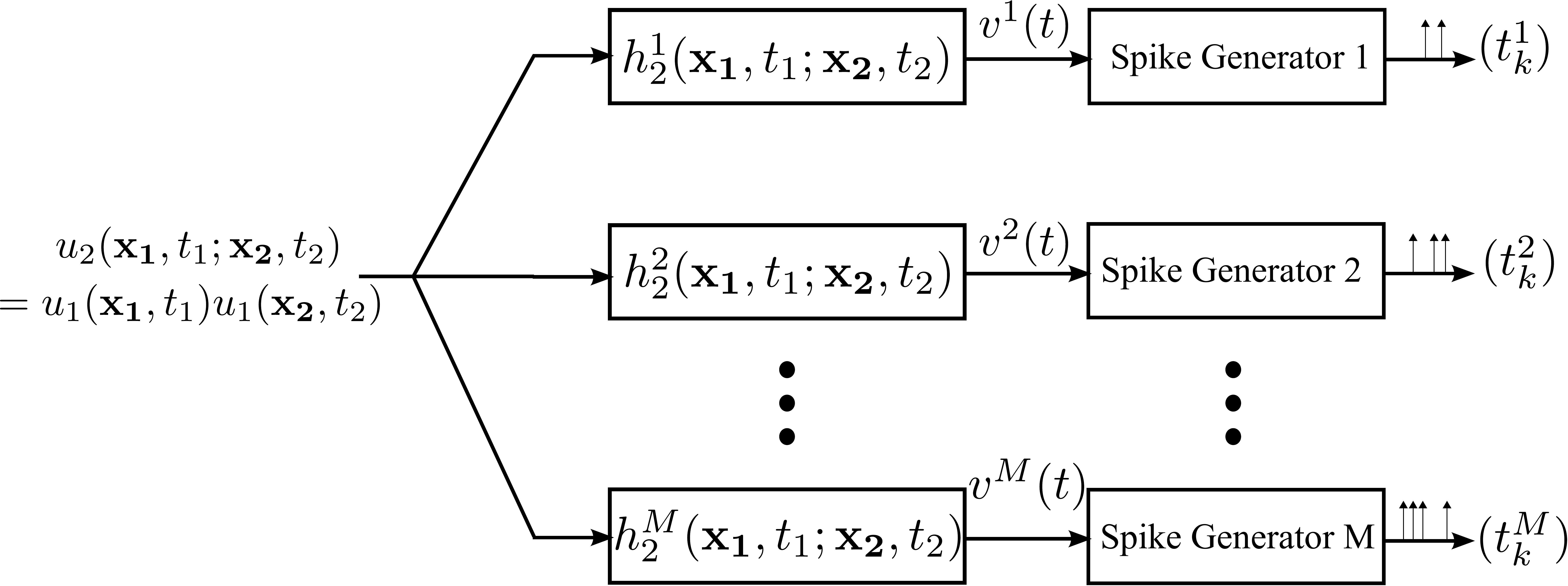}}  ~~
\subfloat[]{ \label{fig:dual_st_id}  \includegraphics[width=0.50\textwidth]{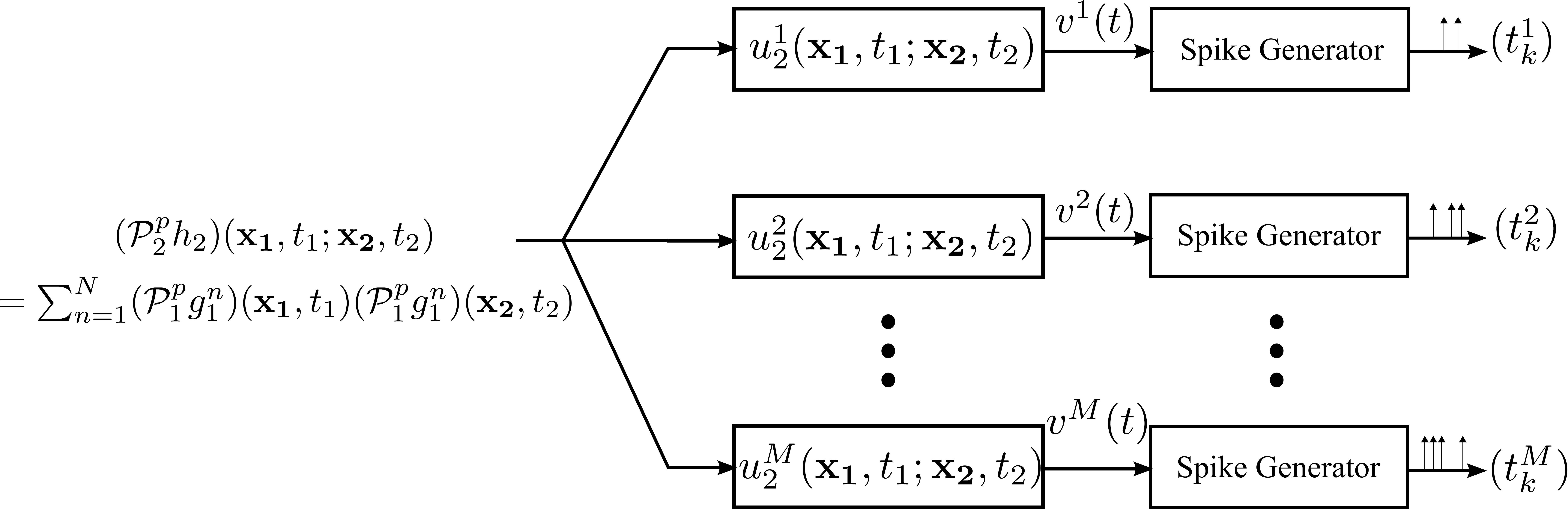}}  ~~
   \caption{Duality between (a) decoding of spatio-temporal stimuli encoded by a neural circuit of $M$ complex cells and
   (b) functional identification of spatio-temporal complex cells by presenting $M$ trials of stimuli.}
   \label{fig:duality_st}
\end{figure}

Similar to the temporal case, the neural circuit is equivalent to that shown in Figure~\ref{fig:dual_st_encode}.
Here, the output of the DSP for each neuron $i=1,2,\cdots,M$, can be expressed as  
\begin{equation}
v^i(t) = \int_{\mathbb{D}^2} h^i_2(\mathbf{x_1},t-s_1;\mathbf{x_2},t-s_2) u_1(\mathbf{x_1},s_1) u_1(\mathbf{x_2},s_2)\mathbf{dx_1}\mathbf{dx_2} ds_1 ds_2.
\label{eq:volterra_output_3}
\end{equation}
Here 
\begin{equation}
h^i_2(\mathbf{x_1},t_1;\mathbf{x_2},t_2) = g^{i1}_1(\mathbf{x}_1, t_1)g^{i1}_1(\mathbf{x}_2, t_2) + g^{i2}_1(\mathbf{x}_1, t_1)g^{i2}_1(\mathbf{x}_2, t_2)
\label{eq:st_h2_form}
\end{equation}
is the low-rank DSP \cite{LAS15}. 
The encoding of stimulus by the neural circuit with complex cells 
is a special case of the low-rank DSP
of the form given in \eqref{eq:st_h2_form}.
When using IAF point neurons as models of the BSGs, we have the following theorem describing the encoding of stimuli.

\begin{lemma}
The encoding of stimulus $u_1\in \hilbert^p_1$ into the spike train sequence
$(t_k^i), k\in\mathbb{I}^i, i=1,2,...,M,$
by a neural circuit of spatio-temporal complex cells is given in functional form by
\begin{equation}
\mathcal{T}^i_k u_2 = q^i_k, k\in\mathbb{I}^i, i = 1,\cdots, M,
\label{eq:t-trans_3}
\end{equation}
where 
$\mathcal{T}^i_k: \hilbert_2^p \rightarrow \real$,
are bounded linear functionals defined by
\begin{equation}
\mathcal{T}^i_k u_2 = \int_{t^i_k}^{t^i_{k+1}} \int_{\mathbb{D}^2} h^i_2 (\mathbf{x_1},t-s_1;\mathbf{x_2},t-s_2) u_2(\mathbf{x_1},s_1;\mathbf{x_2}, s_2) \mathbf{dx_1}\mathbf{dx_2}ds_1 ds_2 dt,
\label{eq:operatorT_3}
\end{equation}
with 
$u_2(\mathbf{x_1},t_1;\mathbf{x_2},t_2) = u_1(\mathbf{x_1},t_1)u_1(\mathbf{x_2},t_2)$.
Finally, $q^i_k = \kappa^i\delta^i - b^i (t^i_{k+1}-t^i_{k})$.
\label{lemma:linear-t-transform_3}
\end{lemma}
\proof As in Lemma \ref{lemma:linear-t-transform}, the t-transform of the $i$-th IAF neuron
is given by \eqref{eq:iaf}.

The relationship \eqref{eq:t-trans_3} follows after replacing $v^i(t)$ given in  \eqref{eq:volterra_output_3} in equation \eqref{eq:iaf}. \qed

Similar to Remark~\ref{rem:gen_sampling}, equation \eqref{eq:t-trans_3} shows that
the encoding of a stimuli by the neural circuit
with low-rank DSPs
can be viewed as generalized sampling.

By abuse of notation, we denote by $\mathbf{c}$
the vector representing the coefficients of $u_1$ in \eqref{eq:uform_3},
and $\mathbf{D}$ as the matrix representing the coefficients of $u_2$ in \eqref{eq:u2form_3}.
We skip here the detailed entries of $\mathbf{c}$ and $\mathbf{D}$ due to the complexity of the indices, but
their construction follows closely with \eqref{eq:c_vector_form} and \eqref{eq:D_matrix_form}, respectively,
and $\mathbf{D} = \mathbf{c}\mathbf{c}^H$.
\begin{theorem}
Encoding the stimulus $u_1\in\hilbert^p_1$
with the neural circuit with complex cells given in \eqref{eq:volterra_output_3} into the spike train sequence
$(t_k^i), k\in\mathbb{I}^i$, $i=1,2,...,M$, satisfies the set of equations
\begin{equation}
\mbox{\bf Tr}( \mbox{\boldmath$\Phi$}^i_k \mathbf{D}) = q^i_k, k\in\mathbb{I}^i, i = 1,\cdots, M,
\label{eq:trace_3}
\end{equation}
where $\mathbf{D} = \mathbf{c}\mathbf{c}^H$ is a rank-$1$ Hermitian matrix and
$(\mbox{\boldmath$\Phi$}^i_k)$,
$k \in \mathbb{I}^i, i = 1,\cdots, M$, are Hermitian matrices.
$[\mbox{\boldmath$\Phi$}^i_k]_{\mathbf{l_{x_2}}l_{t_2}; \mathbf{l_{x_1}}l_{t_1}}$ denotes the entry at the\\
$\left( (l_{t_2}+L_{t_2}+1)\prod_{i=1}^{p}(L_{x_{i2}}+1)+ \sum_{j=1}^{p}(l_{x_{j2}}+L_{x_{j2}} +1) \prod_{i=1}^{j-1}(2L_{x_{i2}}+1) \right)$-th row and the\\
$\left( (l_{t_1}+L_{t_1}+1)\prod_{i=1}^{p}(L_{x_{i1}}+1)+ \sum_{j=1}^{p}(l_{x_{j1}}+L_{x_{j1}} +1) \prod_{i=1}^{j-1}(2L_{x_{i1}}+1) \right)$-th column, and
\begin{equation}
\begin{split}
&[\mbox{\boldmath$\Phi$}^i_k]_{\mathbf{l_{x_2}}l_{t_2}; \mathbf{l_{x_1}}l_{t_1}}  = \\
& \int_{t^i_k}^{t^i_{k+1}} e_{l_{t_1}-l_{t_2}}(t)dt \int_{\mathbb{D}^2} h^i_2(\mathbf{x_1}, s_1;\mathbf{x_2},s_2) e_{\mathbf{\mathbf{l_{x_1}}}, -l_{t_1}}(\mathbf{x_1}, s_1)e_{-\mathbf{l_{x_2}}, l_{t_2}}(\mathbf{x_2}, s_2) \mathbf{dx_1}ds_1\mathbf{dx_2}ds_2,
\end{split}
\label{eq:phi_3}
\end{equation}
where $\mathbf{l_{x_i}} = (l_{x_{1i}}, l_{x_{2i}}, \cdots, l_{x_{pi}}), i = 1,2$.
\label{th:trace-trans_3}
\end{theorem} 
\proof 
Plugging in the general form of $u_2$ in \eqref{eq:u2form_3} into \eqref{eq:operatorT_3}, 
the left hand side of \eqref{eq:t-trans_3} amounts to
\[
\begin{split}
\sum_{\mathbf{l_{x_1}}} \sum_{l_{t_1}}
\sum_{\mathbf{l_{x_2}}}\sum_{l_{t_2}} d_{\mathbf{l_{x_1}},l_{t_1},-\mathbf{l_{x_2}},-l_{t_2}} 
&\int_{t^i_k}^{t^i_{k+1}}e_{l_{t_1}-l_{t_2}}(t) dt \cdot \\
& \cdot \int_{\mathbb{D}^2} h^i_2(\mathbf{x_1},s_1;x_2,s_2)e_{\mathbf{l_{x_1}},-l_{t_1}}(\mathbf{x_1},s_1)e_{-\mathbf{l_{x_2}},l_{t_2}}(\mathbf{x_2},s_2) \mathbf{dx_1dx_2}ds_1ds_2.
\end{split}
\]
It is easy to verify that the expression above can be written as
\begin{equation}
\sum_{\mathbf{l_{x_1}}} \sum_{l_{t_1}}\sum_{\mathbf{l_{x_2}}}\sum_{l_{t_2}} d_{\mathbf{l_{x_1}},l_{t_1},-\mathbf{l_{x_2}},-l_{t_2}}
[\mbox{\boldmath$\Phi$}^i_k]_{\mathbf{l_{x_2}}l_{t_2}; \mathbf{l_{x_1}}l_{t_1}} = 
\mbox{\bf Tr}( \mbox{\boldmath$\Phi$}^i_k\mathbf{D} ),
\label{eq:gtrace}
\end{equation}
where
the \\
$\left( (l_{t_1}+L_{t_1}+1)\prod_{i=1}^{p}(L_{x_{i1}}+1)+ \sum_{j=1}^{p}(l_{x_{j1}}+L_{x_{j1}} +1) \prod_{i=1}^{j-1}(2L_{x_{i1}}+1) \right)$-th row\\
$\left( (l_{t_2}+L_{t_2}+1)\prod_{i=1}^{p}(L_{x_{i2}}+1)+ \sum_{j=1}^{p}(l_{x_{j2}}+L_{x_{j2}} +1) \prod_{i=1}^{j-1}(2L_{x_{i2}}+1) \right)$-th column
entry of $\mathbf{D}$ amounts to 
$\left[\mathbf{D} \right]_{\mathbf{l_{x_1}}l_{t_1} ;\mathbf{l_{x_2}}l_{t_2}} = d_{\mathbf{l_{x_1}}, l_{t_1},- \mathbf{l_{x_2}}, -l_{t_2}}$.

Since $u_2(\mathbf{x_1},t_1;\mathbf{x_2},t_2) = u_1(\mathbf{x_1},t_1)u_1(\mathbf{x_2},t_2)$ and 
$d_{\mathbf{l_{x_1}}, l_{t_1},- \mathbf{l_{x_2}}, -l_{t_2}} = c_{\mathbf{l_{x_1}}, l_{t_1}} c_{\mathbf{l_{x_2}}, l_{t_2}}^H $, thereby
$\mathbf{D} = \mathbf{c}\mathbf{c}^H$.
We also note that since $h^i_2, i=1,\cdots,M$, are assumed to be real valued, $(\mbox{\boldmath$\Phi$}^i_k), k\in\mathbb{I}^i, i=1,\cdots,M$, are Hermitian. \qed

\subsection{Low-Rank Decoding of Spatio-Temporal Visual Stimuli}
\label{sec:st_dec}

When using an algorithm similar to Algorithm~\ref{al:volterra-tdm} to reconstruct spatio-temporal stimuli
encoded by a neural circuit with complex cells, at least $dim(\hilbert^p_1)\left(dim(\hilbert^p_1)+1 \right)/2$ measurements are required.
In addition, at least $dim(\hilbert^p_1)\left(dim(\hilbert^p_1)+1\right)/ (4L_t+1)$ neurons are required, a number that can become unrealistically high with an increasing dimension of the input space.

With the observation that $\mathbf{D} = \mathbf{cc}^H$ is a rank-one matrix, we can apply  algorithms similar to those described in Section~\ref{sec:t_dec} to recover spatio-temporal stimuli encoded by a population of spiking neurons with low-rank DSPs. For the sake of brevity, we skip the details of
the extended Algorithms \ref{al:rankmin} and \ref{al:altmin}, and in what follows we will provide some examples that demonstrate  that
the decoding of spatio-temporal stimuli is still tractable.

\subsubsection{Example - Decoding of 2D Spatio-Temporal Stimuli}
\label{sec:2D_example}

We first present an example in which $\mathbf{x}$ is one-dimensional, \textit{i.e.}, $\mathbf{x}=x_1$.
In this example, our main focus is to illustrate how the number of spikes affects the reconstruction of stimuli encoded by complex cells.

The neural circuit we consider here consists of 62 direction selective complex cells.
The low-rank DSPs of the complex cells are of the form
\begin{equation}
h_2^i(x_{1}, t_1; x_{2}, t_2) = g^{i1}_1 (x_{1}, t_1) g^{i1}_1(x_{2}, t_2) + g^{i2}_1(x_{1}, t_1) g^{i2}_1(x_{2}, t_2) ,
\end{equation}
where $g^{i1}_1(x_1, t)$ and  $g^{i2}_1(x_1, t)$ are quadrature pairs of spatio-temporal Gabor filters and $i=1,\cdots,M$.
The Gabor filters are constructed from dilations and translations of the mother wavelets on a dyadic grid, where the mother functions
can expressed as
\begin{equation}
g^1_1(x_1,t) = \operatorname{exp}\left(- \left(\frac{x_1^2}{8} + \frac{t^2}{0.001} \right) \right) \operatorname{cos}\left( 1.5x_1+20\pi t \right)
\end{equation}
and
\begin{equation}
g^2_1(x_1,t) = \operatorname{exp}\left(- \left(\frac{x_1^2}{8} + \frac{t^2}{0.001} \right) \right) \operatorname{sin}\left( 1.5x_1+20\pi t \right) .
\end{equation}
The BSG of the complex cells are IAF neurons with bias $b^i = 10$ and integration constant $\kappa = 1$, for $i = 1,\cdots,M$.
These two parameters are kept the same for all stimuli. Different threshold values are chosen for the IAF neurons in order
to vary the total number of spikes in a larger range to evaluate  how many measurements are required for a perfect reconstruction of input stimuli.

The domain of the input space $\hilbert^1_1$ is $\mathbb{D} = [0,32]\times[0,0.4]$ ([a.u.] and [sec], respectively) and 
$L_{x_1} = 6, L_t = 4, \Omega_{x_1} = 0.1875\cdot 2\pi, \Omega_t = 10\cdot 2\pi$ [rad/sec]. Thus, $dim(\hilbert^1_1) = 117$.
Stimuli were randomly generated by choosing the basis coefficients
to be i.i.d. Gaussian random variables.

We tested the encoding of $1,416$ stimuli. Each time, a different number
of spikes was generated.
The reconstruction of stimuli was performed in MATLAB using the extended Algorithm~\ref{al:rankmin}, and
the SDPs were solved using SDPT3 \cite{TTT2003}.

\begin{figure}[htbp] 
   \centering
   \subfloat[]{\label{fig:dec_all}\includegraphics[width=0.55\textwidth]{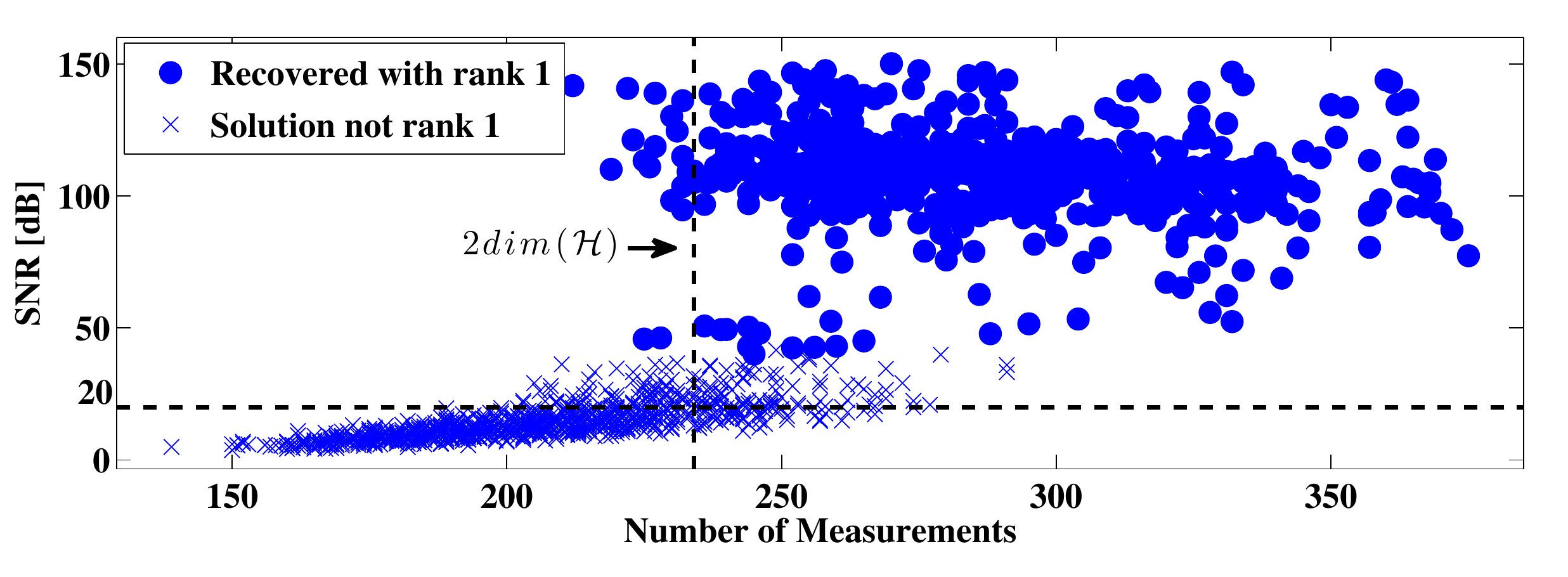}}
   \subfloat[]{\label{fig:dec_rate}\includegraphics[width=0.38\textwidth]{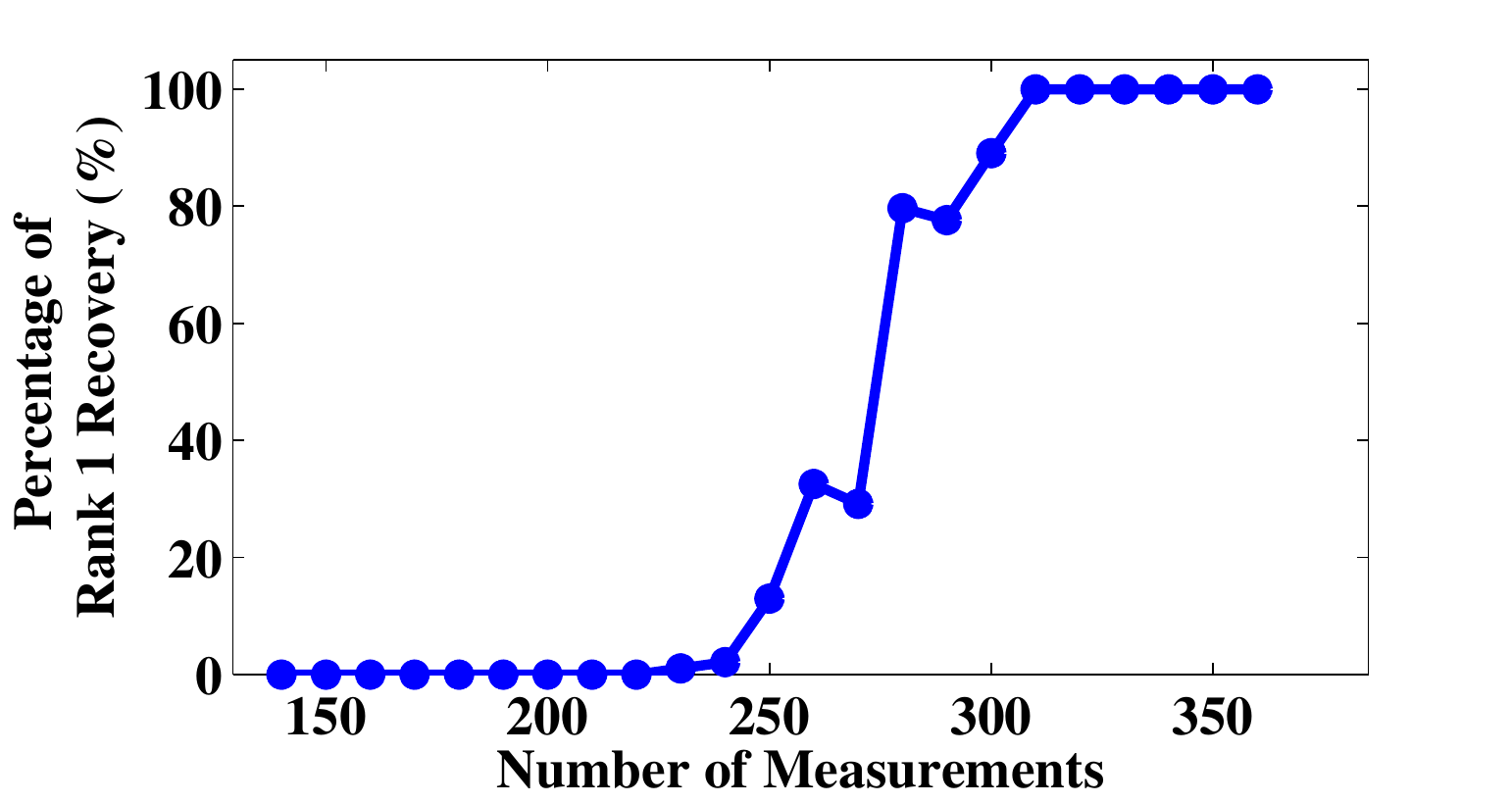}}
   \caption{Example of low-rank decoding of spatio-temporal stimuli. (a) Effect of number of measurements (spikes) on reconstruction quality. (b) Percentage of rank 1 reconstructions.}
   \label{fig:dec_bound}
\end{figure}

The SNR of all reconstructions is depicted in the scatter plot of Figure~\ref{fig:dec_all}.
Here solid dots represent exact rank 1 solutions (largest eigenvalue is at least 100 times larger than the sum of the rest of the
eigenvalues), and crosses indicate that the trace minimization found a higher rank solution with a smaller trace.
The percentage of exact rank 1 solutions is shown in Figure~\ref{fig:dec_rate}. 
Similar to phase transition phenomena in other sparse recovery algorithms
\cite{DMM09}, a relatively sharp transition (around 50 spikes) from very low probability
of recovery to very high probability of perfect reconstruction
can be seen.
It can also be seen that the number of measurements that are needed for perfect
recovery is substantially lower than the $6,965$
spikes required by Algorithm~\ref{al:volterra-tdm}.

\subsubsection{Example - Decoding of 3D Spatio-Temporal Stimuli}
\label{sec:3D_example}

Next, we present two examples of decoding of spatio-temporal visual stimuli encoded
by a population of complex cells.
Here, $\mathbf{x} = (x_1,x_2)$ and
the Volterra DSPs of the complex cells are of the form
\begin{equation}
h_2^i(\mathbf{x_1}, t_1; \mathbf{x_2}, t_2) = g^{i1}_1 (\mathbf{x_1}, t_1) g^{i1}_1(\mathbf{x_2}, t_2) + g^{i2}_1(\mathbf{x_1}, t_1) g^{i2}_1(\mathbf{x_2}, t_2) ,
\end{equation}
where $g^{i1}_1(\mathbf{x}, t)$ and  $g^{i2}_1(\mathbf{x}, t)$ are, for simplicity, quadrature pairs of spatial-only Gabor filters and $i=1,\cdots,M$.
The Gabor filters are constructed from dilations, translations and rotations of a mother wavelets \cite{LPZ2010},
\begin{equation}
g^1_1(\mathbf{x},t) = \operatorname{exp}\left(- \frac{1}{8} \left( 4x_1^2 + 2y_1^2 \right) \right) \operatorname{cos}\left( 2.5x_1 \right)
\end{equation}
and
\begin{equation}
g^2_1(\mathbf{x},t) = \operatorname{exp}\left(- \frac{1}{8} \left( 4x_1^2 + 2y_1^2 \right) \right) \operatorname{sin}\left( 2.5x_1 \right) .
\end{equation}

\begin{figure}[htbp] 
   \centering
   \includemovie[mouse,autoclose,text={\bf\textcolor{red}{(Click to start video)}},inline=\videoinline,controls,toolbar,poster=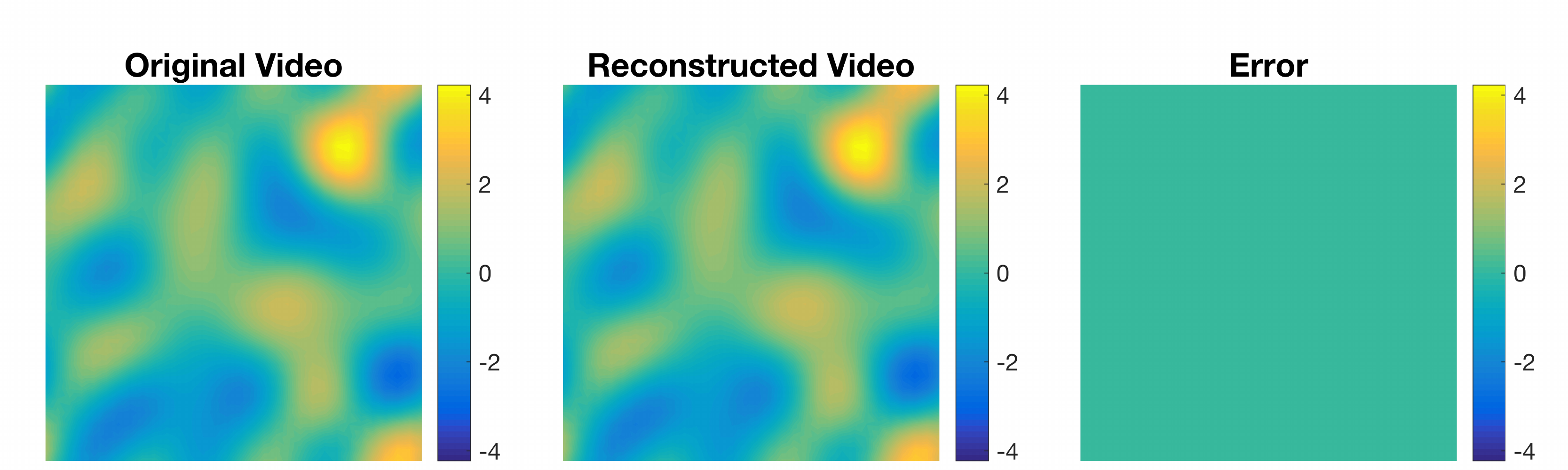]{0.75\linewidth}{0.225\linewidth}{video1.mp4}
      \caption{Example of reconstruction of synthesized visual stimuli.
      Reconstruction of a visual stimulus encoded by $318$ Complex Cells that generated some $1,374$ spikes. A snapshot of the original video is shown on the left. 
      The reconstruction is shown in middle and the error on the right. SNR $92.8$ [dB]. (See also Supplementary Video S1)}
   \label{fig:rec1}
\end{figure}

For the first example, a 0.4-second-long synthetically generated video sequence is encoded by the neural circuit.
The order of the input space  was chosen to be
$L_{x_1} = L_{x_2} = 3, L_t = 4$.
Thus, the dimension of the input space is $441$.
The input stimulus was created by  choosing its basis coefficients
 to be i.i.d. Gaussian random variables.
The stimulus was encoded by a neural circuit consists of $318$ complex cells. 
A total of $1,374$ spikes were generated by the encoding circuit. 
The stimulus was decoded using the extended Algorithm~\ref{al:rankmin}.
As shown in Figure~\ref{fig:rec1}, the video sequence can be perfectly reconstructed with a fairly small number of spikes (A snapshot of the video is shown, see also Supplementary Video S1 for full video).
The SNR of the reconstructed video was $92.8$ [dB], thereby reaching almost perfect reconstruction with machine precision.
Note that without the reconstruction algorithm employed
here, $97,461$ measurements would be required from
at least $5,733$ complex cells to achieve perfect reconstruction.

\begin{figure}[htbp] 
   \centering
   \includemovie[mouse,autoclose,text={\bf\textcolor{red}{(Click to start video)}},inline=\videoinline,controls,toolbar,poster=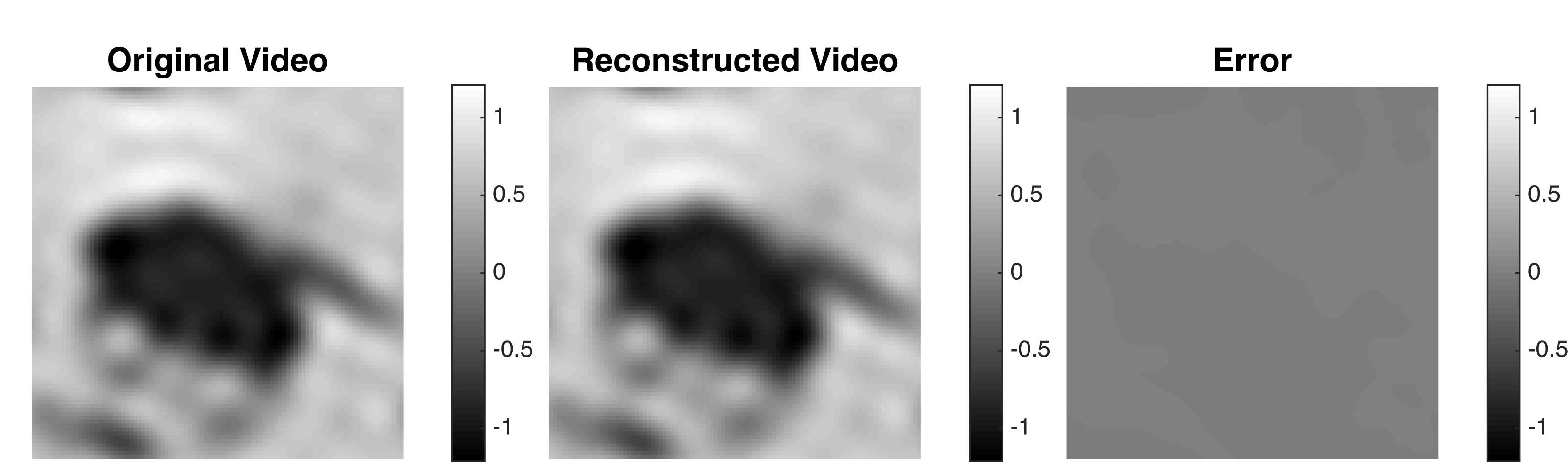]{0.75\linewidth}{0.225\linewidth}{video2.mp4}
      \caption{Example of reconstruction of natural visual stimuli. A natural visual stimulus was encoded by $472$ Complex Cells that generated some $6,000$ spikes. A snapshot of the original video is shown on the left. 
      The reconstruction is shown in middle and error on the right. SNR $68.0$ [dB]. (See also Supplementary Video S2)}
   \label{fig:rec2}
\end{figure}

The second example uses a natural video sequence.
As an illustration, we project the natural video into a space with lower spatial bandwidth
in order to reduce the dimension of the embedding space.
Here, the order of the space is given by $L_{x_1} = 6, L_{x_2} = 6, L_t = 2$, and thus, the dimension of input space is $845$.
A neural circuit consisting of $472$ complex cells was used to encode the stimulus, and a total of $6,000$ spikes were generated.
The stimulus was reconstructed using the extended Algorithm~\ref{al:altmin}.
The number of spikes employed, $\sim 6,000$, is much lower than the  $397,150$ spikes required by Algorithm~\ref{al:volterra-tdm} for perfect recovery.
A snapshot of the original and reconstructed video sequence are shown
in Figure~\ref{fig:rec2} (see also Supplementary Video S2). 
The SNR of the reconstructed video was $68.0$ [dB].

Both examples demonstrate the effectiveness of the reconstruction algorithm proposed in this paper.
In particular, compared to the decoding algorithm based on Theorem~\ref{al:volterra-tdm},
the number of neurons in the neural circuit was substantially reduced.
Therefore, with biologically plausible spike rates and number of complex cells,
the information contained in the spike times
of these neurons can faithfully represent visual stimuli.

\subsection{Low-Rank Functional Identification of Spatio-Temporal Complex Cells}
\label{sec:st_id}

Similar to Section~\ref{sec:identification}, we consider here the identification of low-rank DSP of  complex cells from spike times generated when multiple
stimulus trials are presented. 
We first define the projection operators in $\hilbert^p_1$.
Then, based on \eqref{eq:volterra_output_3}, we show that the duality between
decoding and functional identification also holds in the spatio-temporal case.

\begin{definition}
Let $h_n \in \mathbb{L}^1(\mathbb{D}^n), n=1,2$, where $\mathbb{L}^1$ denotes the space of Lebesgue integrable functions. The operator $\mathcal{P}_{1}^{p}: \mathbb{L}_1(\mathbb{D}) \rightarrow \hilbert^p_1$ given by
\begin{equation}
(\mathcal{P}_{1}^{p} h_1)(\mathbf{x}, t) = \int_{\mathbb{D}} h_1(\mathbf{x}', t') K^p_1(\mathbf{x}, t; \mathbf{x}', t') \mathbf{dx}' dt'
\end{equation}
is called the projection operator from $\mathbb{L}^1(\mathbb{D})$ to $\hilbert^p_1$.
Similarly, the operator $\mathcal{P}_{2}^{p}: \mathbb{L}_1(\mathbb{D}^2) \rightarrow \hilbert_2$ given by
\begin{equation}
(\mathcal{P}_{2}^{p}h_2)(\mathbf{x_1}, t_1; \mathbf{x_2}, t_2) = \int_{\mathbb{D}^2} h_2(\mathbf{x}'_\mathbf{1}, t'_1; \mathbf{x}'_\mathbf{2}, t'_2) K^p_2(\mathbf{x_1}, \mathbf{x_2}, t_1,t_2; \mathbf{x}'_\mathbf{1}, \mathbf{x}'_\mathbf{2}, t'_1,t'_2) \mathbf{dx}'_\mathbf{1} \mathbf{dx}'_\mathbf{2} dt'_1dt'_2
\end{equation}
is called the projection operator from $\mathbb{L}^1(\mathbb{D}^2)$ to $\hilbert_2$.
\end{definition}

We consider here complex cells whose low-rank DSP can be expressed more generally as
\begin{equation}
h_2(\mathbf{x}_1, t_1; \mathbf{x}_2, t_2) = \sum_{n=1}^N g^n_1(\mathbf{x}_1,t_1) g^n_1(\mathbf{x}_2,t_2),
\label{eq:h2_st_form}
\end{equation}
where, by abuse of notation, $g^n_1(\mathbf{x},t), n= 1,\cdots,N$ are impulse responses of spatio-temporal linear filters, and $N \ll dim(\hilbert^p_1)$. Similar to the approach we take in Section~\ref{sec:identification}, this particular structure can be exploited
to identify the projection of $h_2$ using tractable algorithms.

By abuse of notation, we denote $\mathbf{g}^n$ as the vector representing the coefficients of $\mathcal{P}^p_1g^n_1$,
and $\mathbf{H}$ as the matrix representing the coefficients of $\mathcal{P}^p_2h_2$.
The detailed entries of $\mathbf{g}^n$ and $\mathbf{H}$ are constructed similarly to \eqref{eq:pg_vector_form} and \eqref{eq:ph_matrix_form},
respectively.
In addition, we have $\mathbf{H} = \sum_{n=1}^N \mathbf{g}^n(\mathbf{g}^n)^H$.

\begin{theorem} 
By presenting $M$ trials with stimuli $u^i_2(\mathbf{x_1}, t_1; \mathbf{x_2}, t_2) = u^i_1(\mathbf{x_1}, t_1)u^i_1(\mathbf{x_2}, t_2), i=1,\cdots,M$, 
to a complex cell and observing the spike trains $t^i_k, k\in\mathbb{I}^i, i=1,2,\cdots,M$, the coefficients of the projections $\mathcal{P}_{2}^{p}h_2$ of the DSP
of the complex cell, satisfy the set of equations
\begin{equation}
\mbox{\bf Tr}( \mbox{\boldmath$\Psi$}^i_k \mathbf{H}) = q^i_k, k\in\mathbb{I}^i, i = 1,\cdots, M,
\label{eq:trace-st-id}
\end{equation}
where $\mathbf{H}$ is a rank-$N$ positive semidefinite Hermitian matrix and 
$(\mbox{\boldmath$\Psi$}^i_k)$,
$k \in \mathbb{I}^i, i = 1,\cdots, M$, are Hermitian matrices with
the entry at the\\
$\left( (l_{t_2}+L_{t_2}+1)\prod_{i=1}^{p}(L_{x_{i2}}+1)+ \sum_{j=1}^{p}(l_{x_{j2}}+L_{x_{j2}} +1) \prod_{i=1}^{j-1}(2L_{x_{i2}}+1) \right)$-th row and the\\
$\left( (l_{t_1}+L_{t_1}+1)\prod_{i=1}^{p}(L_{x_{i1}}+1)+ \sum_{j=1}^{p}(l_{x_{j1}}+L_{x_{j1}} +1) \prod_{i=1}^{j-1}(2L_{x_{i1}}+1) \right)$-th column 
given by
$[\mbox{\boldmath$\Psi$}^i_k]_{\mathbf{l_{x_2}}l_{t_2}; \mathbf{l_{x_1}}l_{t_1}} =$
\begin{equation}
\int_{t^i_k}^{t^i_{k+1}} e_{l_{t_1},-l_{t_2}}(t)dt \int_{\mathbb{D}^2} u^i_2(\mathbf{x_1}, s_1;\mathbf{x_2},s_2) e_{\mathbf{\mathbf{l_{x_1}}}, -l_{t_1}}(\mathbf{x_1}, s_1)e_{-\mathbf{l_{x_2}}, l_{t_2}}(\mathbf{x_2}, s_2) \mathbf{dx_1}ds_1\mathbf{dx_2}ds_2,
\end{equation}
where $\mathbf{l_{x_i}} = (l_{x_{1i}}, l_{x_{2i}}, \cdots, l_{x_{pi}}), i = 1,2$.
\label{th:dual-st}
\end{theorem} 
Proof: Essentially similar to the proof of Theorem~\ref{th:dual-t}.

\begin{remark}
Theorem~\ref{th:trace-trans_3} and Theorem~\ref{th:dual-st} suggest that decoding of
spatio-temporal stimuli encoded by a population of complex
cells is dual to the functional identification of the DSP of 
complex cells presented with multiple stimulus trials.
This is further illustrated in Figure~\ref{fig:duality_st}.
Note that in identification, only the projection of the complex cell DSP onto the stimulus space can be identified.
\end{remark}

Based on Theorem~\ref{th:dual-st}, we can provide functional identification algorithms for
complex cell DSPs of the form \eqref{eq:h2_st_form}
with a significant reduction in the number of required trials and spikes.
The algorithms are similar to those presented in Section~\ref{sec:t_id}.
In what follows we present a few example of identification of DSPs of complex cells.

\subsubsection{Example - Low-Rank Functional Identification of Complex Cell DSP from Spike Times in Response to Spatio-Temporal Stimuli}
\label{sec:rank_st_id_ex}

In this example, we first consider identifying the DSP of a single complex cell in the neural circuit used 
in Section~\ref{sec:2D_example}.
As a reminder, the neural circuit used in the example in Section~\ref{sec:2D_example} encodes spatio-temporal stimuli
of the form $u_1(x_1,t)$.

We presented to the population of  $M$ complex cells 0.4-second  stimuli,
where $M$ varied from $40$ to $80$.
The stimuli were generated by choosing their basis coefficients as i.i.d. Gaussian random variables.
For each $M$, we repeated the functional identification process for $200$ times, each with different stimuli.
Identification was essentially based on the extended Algorithm~\ref{al:rankmin}, where the SDPs were again solved by SDPT3.

\begin{figure}[h] 
   \centering
  \subfloat[]{\label{fig:id_ex_rate2}\includegraphics[width=0.4\textwidth]{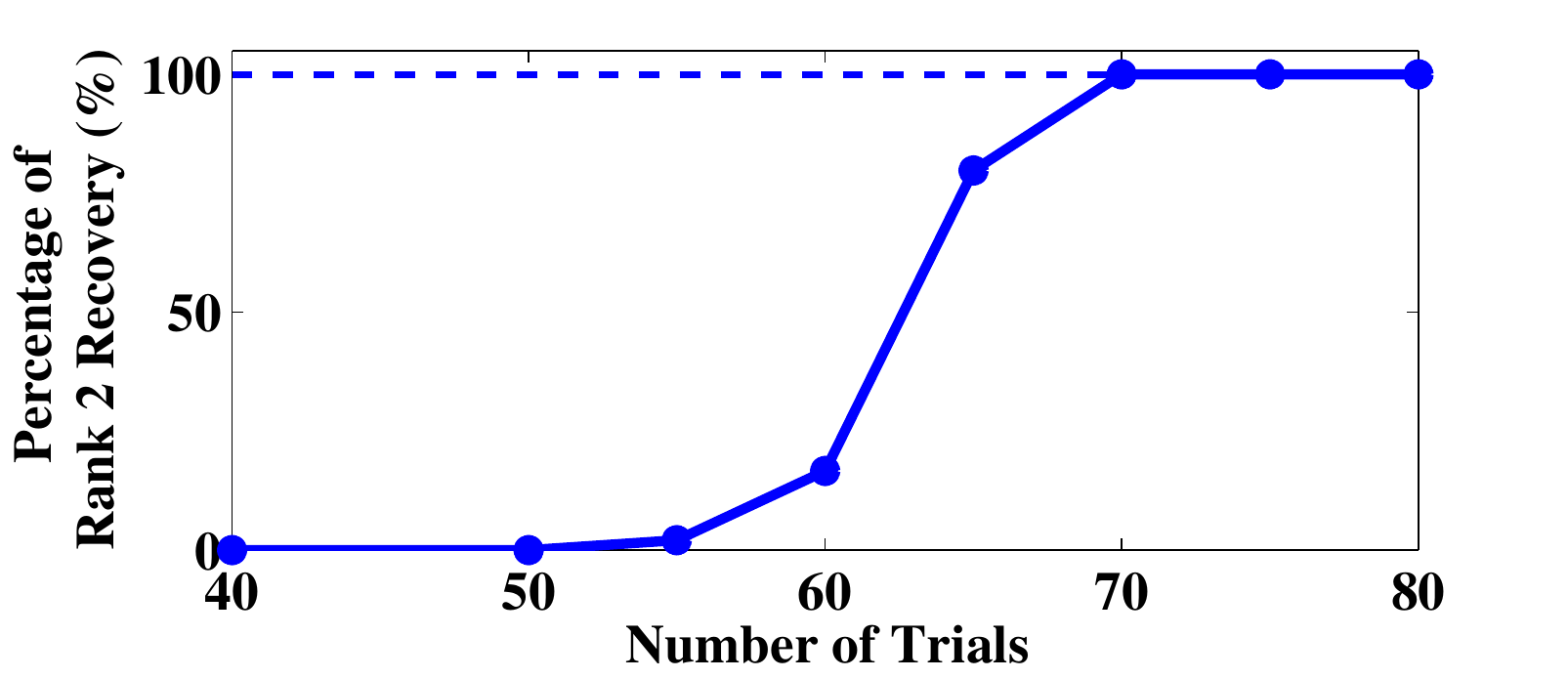}}~~~
  \subfloat[]{\label{fig:id_ex_snr}\includegraphics[width=0.4\textwidth]{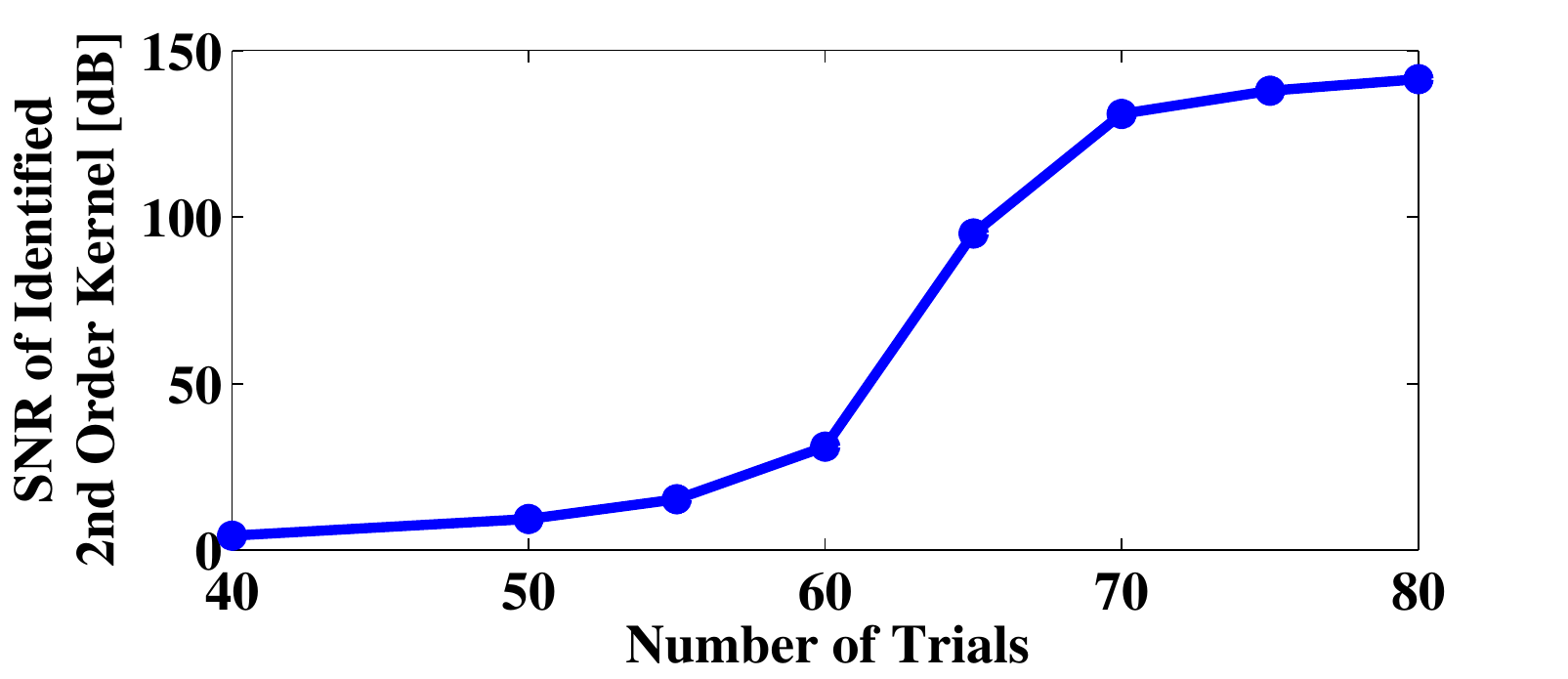}} 
   \caption{Example of low-rank functional identification of spatio-temporal complex cells. (a) Percentage of successful rank 2 recovery in identification. (b) Mean SNR of identified second order DSP kernel.
}
\label{fig:id_ex}
\end{figure}

The percentage of rank 2 solutions is shown in Figure~\ref{fig:id_ex_rate2} as a function of number of experimental trials.
The mean SNR is shown in Figure~\ref{fig:id_ex_snr}.
Figure~\ref{fig:id_ex_rate2} suggests that, if the number of trials is larger than $70$,  
the solution to the trace minimization coincides with high probability with the rank minimization problem. 
In contrast, identification of the complex cell DSP using Algorithm~\ref{al:volterra-cim} would have required at least
$407$ trials.

It can be easily seen that the identification process does not
require a large number of trials to achieve
perfect identification, thereby enabling the identification of non-linear dendritic processing of cells similar in structure to complex cells with a tractable amount of physiological recordings.

\subsubsection{Example - Evaluation of Functional Identification of Neural Circuit of Complex Cells Using Decoding}

We then performed the functional identification of all $62$ complex cells in the neural circuit used of the example in Section~\ref{sec:2D_example}.
Here, our goal is to evaluate the identification quality using decoding.

\begin{figure}[h] 
   \centering
	\includegraphics[width=0.8\textwidth]{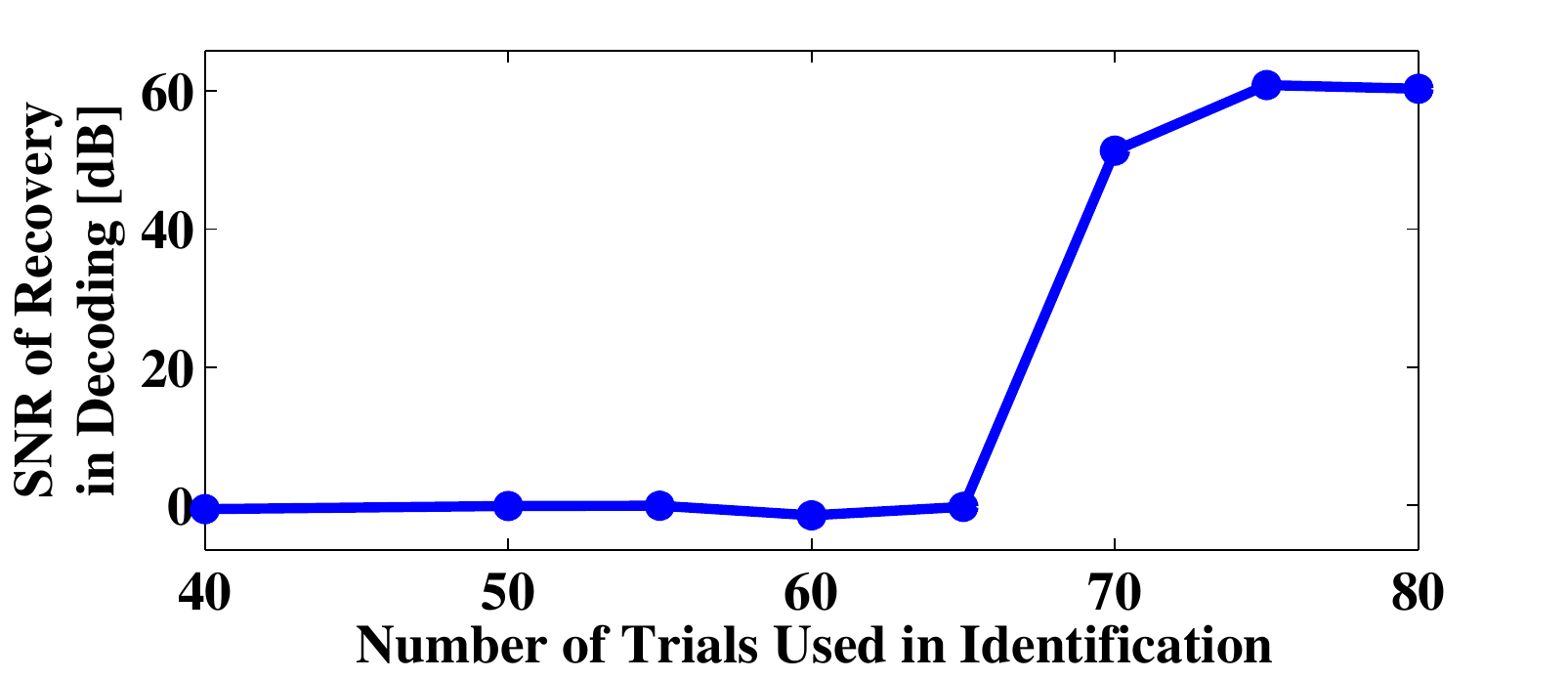}
   \caption{  Evaluating identification quality in the input space. SNR of reconstruction of novel stimuli
assumed to be encoded with the identified DSPs.
}
\label{fig:id_decode}
\end{figure}

We first identified all complex cells by presenting to the neural circuit $M$ spatio-temporal stimuli.
We also performed the identification of the entire circuit using $8$ different values of $M$.
We then presented to the same circuit 100 novel stimuli drawn from the input space and used the spike times
generated by the neural circuit to decode the stimuli.
In the decoding process, we assumed that the DSPs
of the set of complex cells are as identified, for all $8$ values of $M$.
The mean reconstruction SNR of the $100$ stimuli is shown in Figure~\ref{fig:id_decode}.
As shown, the quality of reconstruction was kept at low SNR until enough trials
were used in identification.
When  more than 70 trials were performed, perfect reconstruction was achieved, and thereby the entire neural circuit has been identified with a very high quality.

\subsubsection{Comparison with STC, GQM and NIM}

\begin{figure}[h] 
   \centering
\subfloat[]{\label{fig:stc_compare_curve} \includegraphics[width=0.45\textwidth]{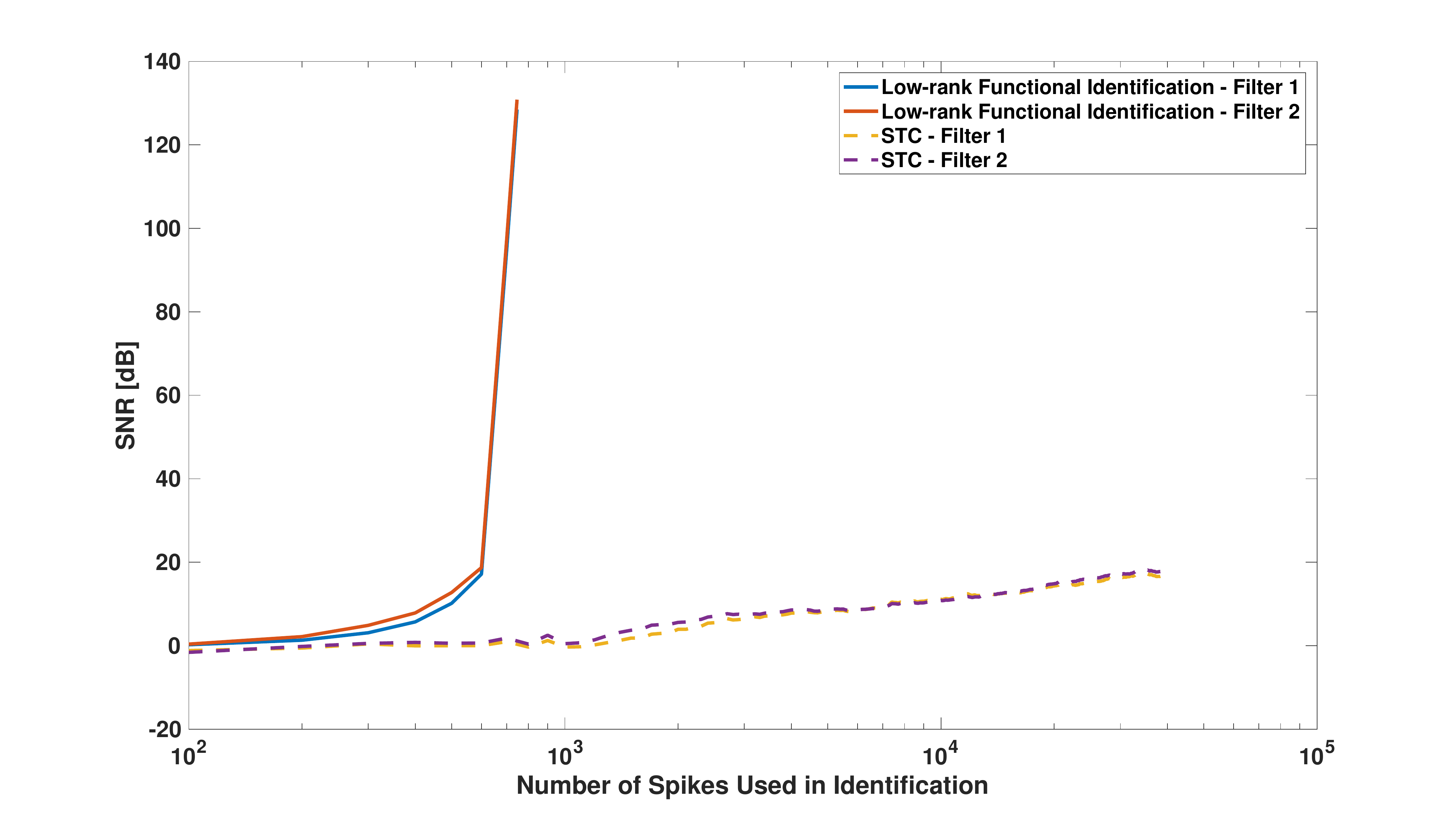}}~
\subfloat[]{\label{fig:stc_compare} \includegraphics[width=0.5\textwidth]{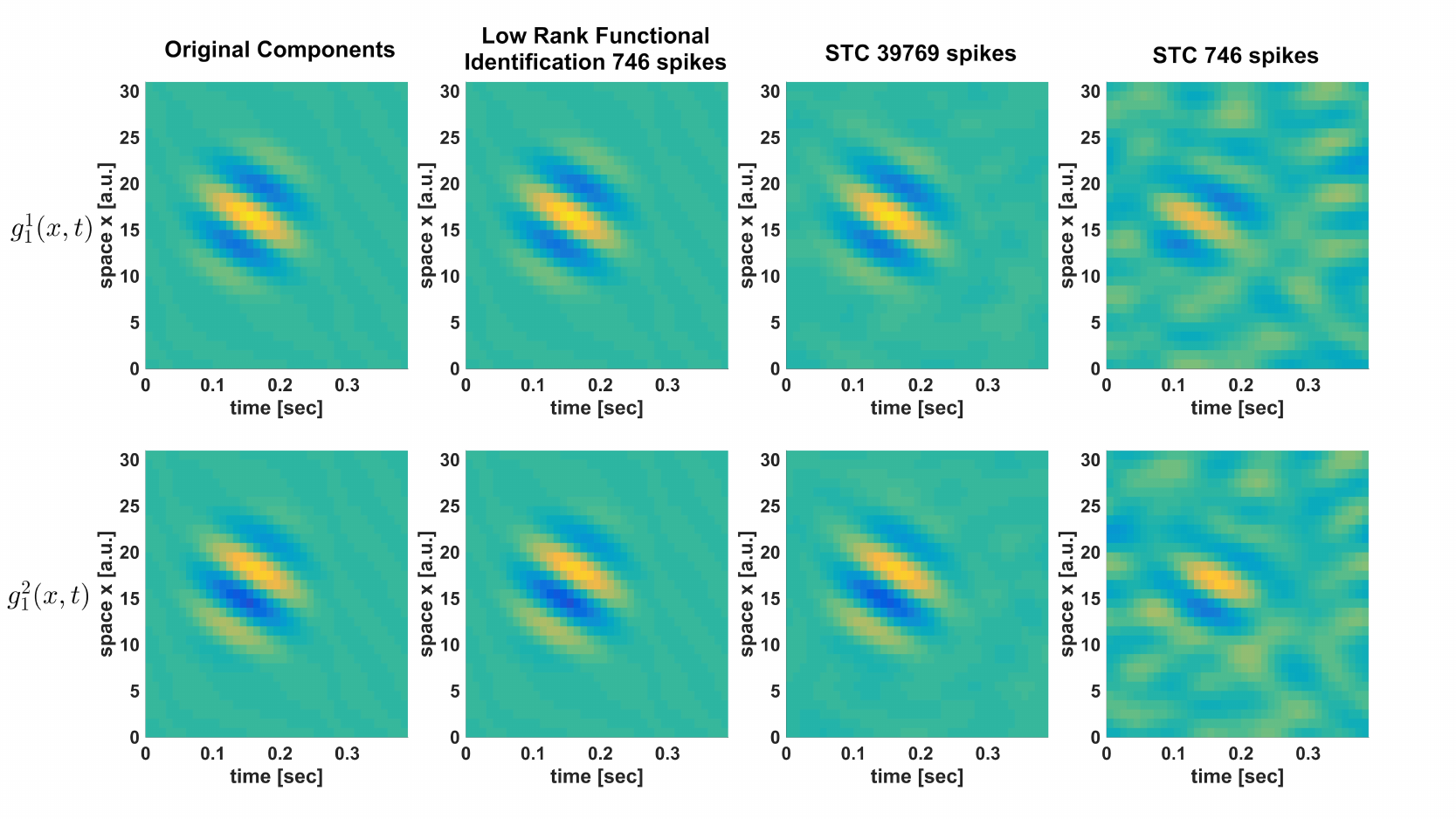}} 
   \caption{ 
Comparison of the low-rank functional identification with STC. (a) SNR of identified quadrature pairs of Gabor filters in a complex cell, as a function of number of spikes used in identification. Low-rank functional identification reaches nearly machine precision with about 746 spikes, which corresponds to about 70 stimulus trials (see also Figure~\ref{fig:id_ex}). STC reaches about 17 [dB] SNR with $\sim30,000$ spikes. (b) Quadrature pair Gabor filters (1st column) identified with low-rank functional identification algorithm with 746 spikes (2nd column, SNR: $128.48$ [dB], $130.84$ [dB]), and with STC using $39,769$ spikes (3rd column, SNR: $16.79$ [dB], $17.88$ [dB])
and using 746 spikes
(4th column, SNR: $0.20$ [dB], $0.60$ [dB]). 
}
\label{fig:comparison}
\end{figure}

We compared the performance of the low-rank functional identification algorithm introduced here
with the widely used Spike-Triggered Covariance (STC) algorithm \cite{PS2006}.
As in Section~\ref{sec:rank_st_id_ex}, a complex
cell with a pair of orthogonal
Gabor filters was chosen for identification.
However, the filters had different norms.

Figure~\ref{fig:stc_compare_curve} shows the quality of identification (SNR) as the number of
spikes used in identification increases.
Note that the low-rank functional identification algorithm reached perfect 
identification using only 746 spikes, whereas the
performance of the
STC algorithm saturated at $\sim17$ [dB] after
almost $40,000$ spikes were used.
Figure~\ref{fig:stc_compare} shows the identified individual Gabor filters of the complex cells
using both algorithms.
The number of spikes used are indicated at the top of each
column.

We also evaluated the identification performance of the generalized 
quadratic model (GQM) \cite{park2013} and the non-linear input model (NIM) \cite{mcfarland2013} with quadratic upstream filters to the same example above.
The results (not shown) were similar to those obtained with the STC algorithm.
 
We note that while the low-rank functional identification algorithm is formulated as non-linear sampling using
TEMs and solved using recent advances in low-rank matrix sensing, the other algorithms tested here rely on moment based or 
likelihood based methods that require a large number of samples to converge.

\section{Conclusions}
\label{sec:conc}

In this paper, we presented sparse algorithms for the reconstruction of temporal as well as spatio-temporal stimuli
from spike times generated by neural circuits consisting of complex cells.
We developed these algorithms by exploiting
the structure of complex cells with  low-rank DSP kernels
and shown that the reconstruction algorithms become
tractable. 
For neural circuits consisting of complex cells, this suggests that, in addition to each extracting visual features, a biologically plausible number of complex cells are capable of faithfully representing visual stimuli.

Based on duality between sparse decoding and functional identification, we showed that functional identification of
complex cells DSPs can be efficiently achieved using similar algorithms as used in decoding. These algorithms makes the functional identification of complex cells
tractable, allowing guaranteed high quality identification using a much smaller set of testing stimuli as well as of shorter time duration.

The mathematical treatment presented here, however, is not limited to the complex cells
in V1.
It can be applied to other neural circuits of interest.
For example, early olfactory coding in 
fruit flies \cite{KLS11} and auditory encoding in grasshoppers \cite{CWR12}
have also been shown to have the structure of low-rank DSP kernels. 
Moreover, the Hassenstein-Reichardt detector \cite{HR56}, a popular model for elementary 
motion detectors in fruit flies, is also I/O equivalent to low-rank
DSP kernels.

\subsubsection*{Competing Interests}
The authors declare that they have no competing interests.

\subsubsection*{Acknowledgments}
The research reported here was supported by AFOSR under grant \#FA9550-16-1-0410.

\newpage
\bibliographystyle{unsrt}
\bibliography{reference}

\end{document}